\documentstyle[aps,epsf,header]{revtex}

\begin{document}
\draft
\title
       {
	The Cauchy Horizon In Black Hole-de Sitter Spacetimes
       }
\author
       {
	Chris M. Chambers
	\thanks{Electronic Address: chris@peloton.physics.montana.edu}
       }
\address
       {
        Department of Physics,
        Montana State University,
        Bozeman,
        MT 59717-3840,
        USA
       }

\maketitle
										   
%
%

\begin{abstract}
The last seven years has produced a growing body of
evidence which concludes that the Cauchy horizon in
black hole de Sitter spacetimes is classically
stable when the surface gravity at the cosmological 
event horizon is greater than that at the Cauchy horizon.
That stability persists for a finite, but non-zero, region of
the black hole's parameter space, $(M,Q,J,\Lambda)$, suggests
that black holes immersed in de Sitter space are counter-examples
to the strong cosmic censorship hypothesis.

In this review we chronicle that body of evidence and
describe the first steps of a program of numerical
work aimed at better understanding the interior  
of black hole-de Sitter spacetimes. The review ends
with a speculative account of the role that future
work will take.

\end{abstract}


\date{\today}
\tighten
\tableofcontents

\vfil
\eject


%
%

\section{Introduction}
\label{cmc-s10}

Our discussions at Haifa have, among other things,
emphasized the need for a 
correct formulation and proof of the {\it Strong
Cosmic Censorship Hypothesis}. Stated in its
simplest, physical form, the strong cosmic
censorship hypothesis pronounces; 
	\begin{quote}
	  All physically reasonable spacetimes are
	  globally hyperbolic -- Apart from
	  a possible initial singularity, no
	  spacetime singularity is ever visible
	  to any observer.
	\end{quote}
Unfortunately cosmic censorship is an area which
yields little return on much effort. We believe,
however, we shall eventually be guided to a
precise formulation and proof by examining 
example and counter-example to the hypothesis.
While counter-examples are abundant in the 
literature, few can live up the claim of being 
`reasonable' spacetimes. There are a few notable
exceptions to this rule though, in particular
the Ernst spacetime, a solution of the
Einstein-Maxwell equations~\cite{cmc-b01} and
the focus of this contribution, black holes
immersed in de Sitter space.   The familiarity
of black hole solutions make the de Sitter
example a particularly attractive field of study.

\subsection*{Black Holes In de Sitter Space}

The fate of an observer falling in to a black hole
is both an interesting and important problem
within the framework of classical general relativity.
The issue takes on a more notable significance if the black
hole is charged or rotating. In this case the
observer's journey appears to continue through the interior
of the black hole and beyond, eventually emerging from a
white hole in to a new
universe. The simplicity of this picture, however,
belies the underlying physics. Deep within the black hole,
concealed from the exterior by an event horizon, 
resides a spacetime singularity, characterized by 
infinite curvatures, and more importantly for our
observer, infinite tidal forces. While fully able
to avoid the crushing grip of the singularity, due in
part to its timelike nature, the observer is unable
to prevent the blatant violation of strong cosmic 
censorship that such a journey irrevocably
promises -- a sighting of the singularity. 
If the issues concerning how physics is going to describe
such a {\it naked} singularity weren't enough,
our observer is faced with an even more fundamental,
though related, problem -- a loss of predictability.
It is well known
that general relativity admits a well posed initial
value problem, i.e., given some suitable initial data on
a spacelike hypersurface, $S$ say, the solution
to the Einstein equations is uniquely determined everywhere 
within the domain of dependence $D(S)$ of $S$~\cite{cmc-b02}.
For some spacetimes though, $D(S)$ can often fail to
cover the entire manifold $M$. So,  even with
suitable initial data on $S$, general relativity is unable
to forecast the evolution of the spacetime beyond
$D(S)$ in these cases. The boundary of $D(S)$, $H(S)$, is called
the Cauchy horizon and marks the division between the 
region where general relativity is able to predict the
evolution and the region where predictability of the field
equations is lost.
Black holes with charge or rotation, such as the
Reissner-Nordstr\"{o}m and Kerr solutions, are well known to 
possess such an horizon. For these black holes the Cauchy
horizon coincides with an inner horizon that veils the
singularity.  In crossing the event horizon, all observer have
unwittingly committed themselves to a journey that will inevitably
encounter and attempt to cross this curious frontier that is
the Cauchy horizon -- an expedition that must contend with naked
singularities, loss of predictability and, in some cases, 
causality violation.
It may come as some relief to hear that physics 
(or nature if you prefer) appears to abhor this situation too
and, indeed, conspires to prevent the passage through the interior,
terminating it at the Cauchy horizon~\cite{cmc-b03}.
The conspiracy itself is rooted in an instability of the inner horizon to
even the smallest time dependent perturbation, converting an initially
regular horizon in to a null, spacetime curvature singularity, 
effectively sealing off the tunnel to other worlds.
The source of the instability can be understood quite simply
as an {\it infinite proper time compression} effect. Because
the causal past of the Cauchy horizon contains the entire 
universe external to the black hole, any observer approaching the
horizon sees an infinite number of events in a finite proper time.
Stated more dramatically; the observer sees the entire history 
of the external universe flash before their eyes,
in the last few moments before they cross the Cauchy horizon. 
That this compression effect becomes larger as one approaches
the Cauchy horizon, results in progressively larger energy densities being
measured,  until, at the horizon itself, they become infinite,
and turn it in to a spacetime singularity.
Thus the Cauchy horizon instability restores predictability
to the situation and provides an excellent example of a 
strong cosmic censorship obeying spacetime.
So what if the causal past of the Cauchy horizon does not
contain the entire external universe but just some
part of that universe? In that case one
can easily conceive a scenario in which the observer crossing
the Cauchy horizon sees only a finite number of events
in a finite proper time, possibly leading to
no infinite blueshift effects and hence no infinite energy densities 
or curvature singularities at the horizon. In effect, such
a spacetime could have a Cauchy horizon that is stable
to time dependent perturbations and serve as a
counter-example to strong cosmic censorship (reasonable
or unreasonable). Does nature, in an attempt to prevent
the embarrassment of naked singularities, plot to prevent such
a spacetime occurring in a physically realistic situation? 
Amazingly it appears not. As we mentioned earlier,
black holes immersed in de Sitter space prove to be notable
counter-examples, for de Sitter black holes are  part of
a closed universe and play out much of the scenario
discussed above. The familiarity of black hole solutions,
combined with the knowledge that all the known black hole 
spacetimes of the Kerr-Newman family can be generalized to 
include a cosmological constant, establishes black hole de 
Sitter spacetimes as a favorable field of investigation.

During the last seven years a consistent picture of a classically 
stable Cauchy horizon in black hole-de Sitter spacetimes has emerged. 
Both linear studies and non-linear backreaction calculations 
indicate that the Cauchy horizon is stable. Although
the scenario envisaged above, based on the infinite time compression 
effects, suggests the Cauchy horizon will always be stable, 
all analyses agree that stability persists for only a finite, but
non-zero, measure on the parameter space $(M,Q,J,\Lambda)$
of the black hole, infringing on the very spirit of the
strong cosmic censorship hypothesis. 
Whether one views these spacetimes as reasonable 
counter-examples to the strong cosmic censorship hypothesis is a 
matter for personal tastes. In the past, much observational 
evidence was advanced to suggest that the cosmological constant, made
famous by Einstein, is zero. Lately (within the last year),  new
evidence, based on improved gravitational lensing and cosmic microwave 
background observations, has surfaced and is currently challenging 
the established view of a vanishing cosmological constant~\cite{cmc-b04}. 
At the least, since a cosmological constant does not violate any 
known law of physics, black hole-de Sitter spacetimes provide an 
excellent arena in which to examine the current form of the strong cosmic 
censorship hypothesis. The appeal of black hole-de Sitter
spacetimes is also realized when comparing them with their
asymptotically flat counterparts. While the Cauchy horizon
in the Reissner-Nordstr\"{o}m spacetime is well known to
be unstable it is intruiging that the metric, expressed in well
behaved coordinates, is regular there. Further to this, it
has been demonstrated that while the tidal forces, due to the
strong gravitational field of the Cauchy horizon singularity,
on an observer grow without bound as the Cauchy horizon is
approached the actual tidal deformation suffered by an observer
does not grow without bound. These facts have re-opened the
issues of traversability of the Cauchy horizon, and formed a
major part of the discussions at Haifa.  The answers to this
debate are by no means clear. It has been proposed that
the finite tidal deformations are irrelevant to the question 
of traversability~\cite{cmc-b05} -- any infalling object 
is completely destroyed at the Cauchy horizon since the 
energy absorbed by the object diverges as the horizon is approached.  
Our black hole-de Sitter model has none of these interpretational
problems~\cite{cmc-b06}. In this respect, the dilemmas
that a traversible Cauchy horizon faces us with, are more clear
cut in the case of black hole-de Sitter spacetimes than they are in
the asymptotically flat case.

\subsection*{Arrangement}

In accordance with the requirements of the organizers of the
``Internal Structure of Black Holes and Spacetime Singularities"
meeting at Haifa, this contribution has been written at a level and
conciseness specifically intended for those who are conversant with
General Relativity but are not familiar with the subject of Cauchy
horizon stability, 
particularly graduate students for whom the subject literature, and logic 
behind it, is often confusing.  That said, it is also hoped that those 
working in this and related fields will find this a worthy offering 
to the subject area. While every attempt has been made to present an 
impartial account of the subject, the narration undoubtedly suffers
from my own personal bias.

In section~\ref{cmc-s20}  we present a mathematical description of 
black hole-de Sitter spacetimes.
The section attempts to provide most of the basic
mathematical material that is required for an understanding of
later sections. Where this is not been possible the reader has been
guided to compensating reference material.
For the sake of simplicity, attention is focused exclusively on
the spherically symmetric Reissner-Nordstr\"{o}m-de Sitter
solution.  Much of what needs to be known about black hole-de Sitter
spacetimes is most easily gained through studying  the
Reissner-Nordstr\"{o}m-de Sitter spacetime.
The different coordinate systems used throughout the spacetime, and the
role they play, is explained and emphasized with the section concluding
on a study of the behavior of radially free-falling geodesic 
observers. While the purpose of some of the concepts may not 
be immediately apparent, they will be become more clear on a 
second read through.

Section~\ref{cmc-s60} chronicles an account of the relevant 
works performed to date. While each commentary is not exhaustive, 
every attempt has been made to produce a clear and concise summary
of that particular study. Specific emphasis has been placed on the motivation
and philosophy of the approach and a precise statement of the 
results obtained. Each account is followed by critical review
of the method, and results, 
in an endeavor to provide the continuity required to go from 
one study to the next and to illustrate the logic in that step.

In Sec.~\ref{cmc-s130} we provide additional material, in the
form of comments, that have not, hereto, been published. 
Much of this material originates from the questions and
comments that have arisen at previous conferences and from
the many personal discussions that occurred during the Haifa meeting.
With the knowledge of Sec.~\ref{cmc-s60} at hand, these questions
are explained and  an attempt to address them is made.
The conclusions of this section leads us nicely on to the 
next section.

The next section, Sec.~\ref{cmc-s140}, elaborates on the first steps of
an ongoing program of numerical work aimed at understanding the
interior of black hole-de Sitter spacetimes in more detail.
The results and motivation of this first step, an investigation
of the late time behavior of fields during gravitational collapse in
de Sitter space, are presented.
The conclusions of this study are discussed, with particular 
emphasis placed on their compatability with the previous analytic
studies of the interior.

Finally, we conclude in Sec.~\ref{cmc-s160} with a
summary of the results presented in this
contribution. In keeping with the mood of the
Haifa meeting we also take this opportunity to
speculate on the role that future work might take in the study of the
Cauchy horizon in black hole-de Sitter spacetimes.

We also include two appendixes which provide additional
material to the contribution, but need not be read
in conjunction with the main body of text. An exhaustive
bibliography is supplied at the end, which contains additional
comments relating to each section.

\subsection*{Acknowledgments}

It is a pleasure to thank the organizers of 
``The Internal Structure of Black Holes and
Spacetime Singularities" workshop, at Haifa, for
a pleasant meeting and for the generous 
hospitality offered during our stay.
In particular my warm thanks go to Lior Burko,
Amos Ori and Liz Youdim for making our visit
to Haifa an extremely enjoyable one. 
I am also grateful 
for discussions with Alfio Bonnano, Patrick Brady, 
Eanna Flanaghan, Eric Poisson, Ian Moss and Amos 
Ori -- all of which have aided this review. 

Chris M. Chambers is a Fellow of The Royal
Commission for the Exhibition of 1851, who's
financial support is gratefully acknowledged.
This work was supported in part by NSF
Grant No. PHY-9722529 to North Carolina State University,
and NSF Grant No. PHY-9511794 to Montana State University.

\vfil
\eject
%
%

\section{Reissner-Nordstr\"{o}m-de Sitter Black Hole}
\label{cmc-s20}

\subsection{The Spacetime}

\label{cmc-s30}

The generalization of the Reissner-Nordstr\"{o}m black hole
solution to include a cosmological constant has been
provided by Carter~\cite{cmc-b10}. In terms of advanced
Eddington-Finkelstein coordinates $(v,r)$, the
Reissner-Nordstr\"{o}m-de Sitter metric is
	\begin{equation}
	  ds^{2} = -f(r) dv^{2} + 2 dv dr + r^{2}
	  d \Omega^{2} \; ,
	\label{cmc-e10}
	\end{equation}
where
	\begin{equation}
	  f(r) = 1 - \frac{2 M}{r} + \frac{Q^{2}}{r^{2}}
	  - \frac{r^{2}}{\alpha^{2}}
	  \ \ \ , \ \ \
	  \alpha^{2} = \frac{3}{\Lambda} \; .
	\label{cmc-e20}
	\end{equation}
In Eq.~(\ref{cmc-e20}) $M$ is the Bondi mass of the black hole,
$Q$ its electric charge and $\Lambda$ is the cosmological
constant. Throughout what follows, it will be assumed that
$\Lambda$ is a positive, non-zero, constant. The function
$d \Omega^{2} = d \theta^{2} + \sin^{2} (\theta) d \varphi^{2}$
is the metric on the unit two sphere. The coordinate $v$ is
the standard advanced time coordinate, related to the
Schwarzschild time coordinate $t$ by
	\begin{equation}
	  v = t + r_{*} \; ,
	\label{cmc-e30}
	\end{equation}
with
	\begin{equation}
	  r_{*} = \int \frac{dr}{f(r)} = 
	  - \frac{1}{2 \kappa_{1}} \ln \left| \frac{r}{r_{1}}
	  -1 \right|
	  + \frac{1}{2 \kappa_{2}} \ln \left| \frac{r}{r_{2}}
	  -1 \right|
	  - \frac{1}{2 \kappa_{3}} \ln \left| \frac{r}{r_{3}}
	  -1 \right|
	  + \frac{1}{2 \kappa_{4}} \ln \left| \frac{r}{r_{4}}
	  -1 \right| \; ,
	\label{cmc-e40}
	\end{equation}
where the arbitrary constant of integration has, for simplicity,
been set to zero. The constants $\kappa_{j}$ represent the
surface gravity~\cite{cmc-b20} at the corresponding $j^{\rm th}$ horizon,
which are located at $r=r_{j}$. In general, the surface
gravity is given by
	\begin{equation}
	  \kappa_{j} = \lim_{r \rightarrow r_{j}}
	  \sqrt{
	  \frac{(\nabla_{\mu} | {\boldmath \ell}^{2} | )
	  ( \nabla^{\mu} |{\boldmath  \ell}^{2} | )}{ 4 
	  | {\boldmath \ell}^{2} |}
	  } \; ,
	\label{cmc-e50}
	\end{equation}
where ${\boldmath \ell}$ is some suitably normalized Killing vector that
is null on the $j^{\rm th}$ horizon and ${\boldmath \ell^{2}} = \ell_{\mu}
\ell^{\mu}$. In the case of a static, spherically symmetric
spacetime with a line element similar to Eq.~(\ref{cmc-e10})
\cite{cmc-b30}, then Eq.~(\ref{cmc-e50}) reduces to
	\begin{equation}
	  \kappa_{j} = \frac{1}{2} \left| 
	  \frac{d f}{d r} \right|_{r=r_{j}} \; ,
	\label{cmc-e60}
	\end{equation}
which can be verified using Eq.~(\ref{cmc-e10}) and 
${\boldmath \ell } =
\partial_{v}$. The location of the horizons $r_{j}$ is given
by the roots to the equation
	\begin{equation}
	  f(r) \equiv 0 \; ,
	\label{cmc-e70}
	\end{equation}
which can be seen [via Eq.~(\ref{cmc-e20})] to reduce to a
quartic equation in $r$. In general there will be four 
distinct roots~\cite{cmc-b40} which we label $r_{1}, r_{2}, 
r_{3}$ and $r_{4}$. The lack of a cubic term in 
Eq.~(\ref{cmc-e70}) forces at least one of the roots to
be negative and unphysical in this spacetime~\cite{cmc-b50}.
We order the roots as follows
	\begin{equation}
	  r_{1} > r_{2} > r_{3} > r_{4} 
	  \ \ \ {\rm where} \ \ \
	  r_{4} < 0 \; .
	\label{cmc-e80}
	\end{equation}
The root $r_{1}$ denotes the location of the {\it cosmological}
event horizon, $r_{2}$ the location of the {\it black hole} event
horizon and $r_{3}$ the location of the {\it inner} horizon
(which from now on we shall refer to as the {\it Cauchy} horizon)
of the black hole spacetime. We note that although $r_{4}$ does
not correspond to a physical horizon in this spacetime, we
are still able to define the constant $\kappa_{4}$ through
Eq.~(\ref{cmc-e60}). Though not strictly correct, $\kappa_{4}$
is often still referred to as a surface gravity. By factoring
the metric function $f(r)$ as
	\begin{equation}
	  f(r) = - \frac{(r - r_{1}) (r - r_{2})
	  (r - r_{3}) (r - r_{4})}{\alpha^{2} r^{2}} \; ,
	\label{cmc-e90}
	\end{equation}
it is a simple task to show that the surface gravities are
	\[
	  \kappa_{1} = \frac{(r_{1} - r_{2}) (r_{1} - r_{3})
	  (r_{1} - r_{4})}{2 \alpha^{2} r_{1}^{2}}
	  \ \ , \ \
	  \kappa_{2} = \frac{(r_{1} - r_{2}) (r_{2} - r_{3})
	  (r_{2} - r_{4})}{2 \alpha^{2} r^{2}_{2}}
	\]
	\begin{equation}
	  \kappa_{3} = \frac{(r_{1} - r_{3}) (r_{2} - r_{3})
	  (r_{3} - r_{4})}{2 \alpha^{2} r_{3}^{2}}
	  \ \ \ {\rm and} \ \ \
	  \kappa_{4} = \frac{(r_{1} - r_{4}) (r_{2} - r_{4}) 
	  (r_{3} - r_{4})}{2 \alpha^{2} r_{4}^{2}} \; .
	\label{cmc-e100}
	\end{equation}
Using Eqs.~(\ref{cmc-e40}),~(\ref{cmc-e90}) and (\ref{cmc-e100})
it is not difficult to show that
	\begin{equation}
	  \lim_{r \rightarrow r_{j}} f(r) =
	  \pm 2(-1)^{j} r_{j} \kappa_{j} \  e^{2 (-1)^{j} 
	  \kappa_{j} r_{*}} \; ,
	\label{cmc-e110}
	\end{equation}
where $(+)$ indicates that we approach the $j^{\rm th}$ horizon from
above and the $(-)$ from below.

\subsection{Coordinate Systems}

\label{cmc-s40}

%
%
\begin{figure}[h]
\begin{center}
\epsfxsize=0.3\textwidth
{\leavevmode\epsffile{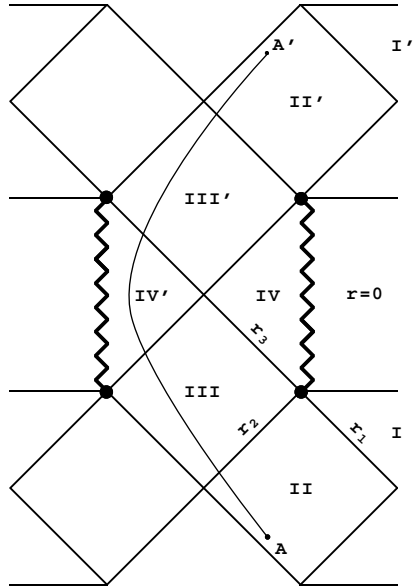}}
\end{center}
\vspace{0.5cm}
\caption{
A portion of the Penrose conformal diagram for the 
Reissner-Nordstr\"{o}m-de Sitter black
hole spacetime. The locations of the
cosmological event horizon at $r=r_{1}$,
the black hole event horizon at $r=r_{2}$
and the inner horizon at $r=r_{3}$, which is also a Cauchy
horizon for external initial value problems, are shown.
The dark wavy line represents the
timelike spacetime singularity at $r=0$.
Also shown are the path of an infalling
observer AA$'$, the four main spacetime
regions I-IV and their isometric counterparts
labelled I$'$-IV$'$. The diagram can be extended
in all direction to reveal an infinite lattice
of asymptotically de Sitter universes.}
\label{cmc-f10}
\end{figure}

The Penrose conformal diagram~\cite{cmc-b60} for the
Reissner-Nordstr\"{o}m-de Sitter black hole spacetime
is shown in Fig.~\ref{cmc-f10}. As well as the location
of the three horizons $r_{1}, r_{2}$ and $r_{3}$, the
timelike spacetime singularity at $r=0$, and the path
of radially infalling observer (see Sec.~{\ref{cmc-s50})
AA$'$ are also shown. It is clear from the conformal
diagram that the horizons divide the spacetime in to
four main regions, labelled I-IV in Fig.~\ref{cmc-f10},
	\[
        \begin{array}{lccc}
	    {\rm Region} & {\rm I} & : & r_{1} < r < \infty   \\
	    {\rm Region} & {\rm II} & : & r_{2} < r < r_{1}   \\
	    {\rm Region} & {\rm III} & : & r_{3} < r < r_{2}  \\
   	    {\rm Region} & {\rm IV} & : & 0 < r < r_{3}.
	\end{array}
	\]
Figure~\ref{cmc-f10} also displays the primed regions I$'$-
IV$'$ which denote those regions of spacetime that are
isometric to regions I-IV. In studying issues related to
the Cauchy horizon, we are predominantly interested in
region II, which we shall frequently refer to as the
exterior, and region III, which we shall refer to as the
interior. In physical terms, we can consider region II
as the {\it birth-site} for field fluctuations such as
electromagnetic and gravitational waves~\cite{cmc-b70}.
These waves are continually scattered off the spacetime
curvature, a fraction of which is transmitted across the
cosmological event horizon in to region I, while the
remainder is scattered through the black hole event horizon
in to region III. The waves crossing the black hole event
horizon eventually propagate throughout the interior
region, perturbing the spacetime geometry there, and in
particular the Cauchy horizon. It will, therefore, be
beneficial if some time is devoted to each of these 
regions, paying special attention to the coordinate
systems used there. For this purpose it proves useful
to write the metric [Eq.~(\ref{cmc-e10})] in 
standard diagonal form using Eq.~(\ref{cmc-e30}),
	\begin{equation}
	  ds^{2} = -f dt^{2} + f^{-1} dr^{2}
	  + r^{2} d \Omega^{2} \; .
	\label{cmc-e120}
	\end{equation}
Figure~\ref{cmc-f20} shows enlarged versions of the regions II,
and III, of Fig.~\ref{cmc-f10}. In each region, surfaces of
constant $r$, the unbroken lines, and surfaces of constant
$t$, broken lines, are shown. The corners of the regions
are labelled $A,B,C,D,E$ and $F$. These points can be
seen to be singular points of the conformal diagram, in 
the sense that there is no one to one mapping between the
corner points and the coordinates $(t,r)$. For the sake of clarity
we shall consider each region individually.

%
%
\begin{figure}[h]
\leavevmode
\begin{center}
\epsfxsize=0.3\textwidth
\leavevmode\epsffile{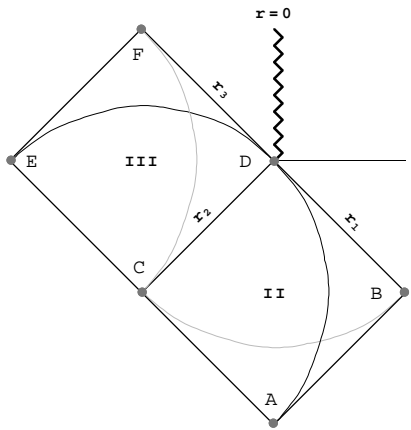}
\end{center}
\vspace{0.5cm}
\caption{The conformal diagram of the exterior and
the interior, regions II and III respectively, of the 
Reissner-Nordstr\"{o}m-de Sitter black hole spacetime. 
The unbroken curved lines represent surfaces of constant $r$ 
whilst the broken curved lines represent surfaces of constant 
$t$. For orientation with {\protect Fig.~\ref{cmc-f10}} 
the horizons an singularity are also indicated.}
\label{cmc-f20}
\end{figure}

\subsection*{Region II}

In region II, depicted in Fig.~\ref{cmc-f30}, we define
two radial, null coordinates $u$ and $v$ by
	\begin{eqnarray}
	  v &=& t + r_{*} 
        \label{cmc-e130} \; ,
	  \\
	  u &=& t - r_{*}
	\label{cmc-e140}
	\end{eqnarray}
where $v$ is defined exactly as in Eq.~(\ref{cmc-e30}). In terms
of $(u,v)$ the spacetime metric [Eq.~(\ref{cmc-e10})] assumes
the double null form
	\begin{equation}
	  ds^{2} = - f du dv + r^{2} d \Omega^{2} \; ,
	\label{cmc-e150}
	\end{equation}
whose form will be useful in Sec.~\ref{cmc-s150}, when we study 
scalar wave propagation  on a Reissner-Nordstr\"{o}m-de Sitter
background. Surfaces of
constant $u$ and $v$ are shown in Fig.~\ref{cmc-f30} as
lines at $45^{o}$ to the horizontal, with $v={\rm const.}$
surfaces running parallel to the (future) cosmological 
event horizon $BD$ and $u={\rm const.}$ surfaces running 
parallel to the (future) black hole event horizon $CD$.
As indicated in the
figure, we have defined the time coordinate $t$ in 
Eq.~(\ref{cmc-e120}) so that it is minus infinity on
$CAB$ and plus infinity on $CDB$. Using Eq.~(\ref{cmc-e40})
one can determine that $r_{*}$ takes the values minus
infinity on $ACD$  and plus infinity on $ABD$. With the
definitions of $u$ and $v$ as above, [Eqs.~(\ref{cmc-e130}) 
and (\ref{cmc-e140})], it is not difficult to see that
$v$ assumes the values minus infinity on the (past)
black hole event horizon $AC$ and plus infinity on
the (future) cosmological event horizon $BD$, whilst
$u$ is minus infinity on the (past) cosmological event
horizon $AB$ and plus infinity on the (future) black hole
event horizon $CD$. In order to analytically continue the
spacetime across the boundaries $AB, BD, CD$ and $AC$
one has to introduce new coordinates that are regular
there (unlike $u$ and $v$ which diverge). The new coordinates
are furnished by the dimensionless Kruskal-Szekeres 
coordinates, conventionally labelled $U$ and $V$. In later
sections we shall be interested in both the (future) 
cosmological  and black hole event horizons. For this reason 
we shall consider the $(U,V)$ coordinates for the boundaries 
$BD$ and $CD$ only, the extension to the other boundaries
being self-evident. Near $CD$ we define
	\begin{eqnarray}
	  U &=& - e^{- \kappa_{2} u} \; ,
	\label{cmc-e160}
	  \\
	  V &=& e^{\kappa_{2} v}
	\label{cmc-e170}
	\end{eqnarray}
so that $U$ tends to zero as $u$ tends to plus infinity. In terms
of these two coordinates, the metric, Eq.~(\ref{cmc-e150}), close
to the (future) black hole event horizon becomes
	\begin{equation}
	  ds^{2} \simeq \frac{2 r_{2}}{\kappa_{2}} dU dV
	  + r^{2} d \Omega^{2} \; ,
	\label{cmc-e180}
	\end{equation}
which is regular at the black hole event horizon. Similarly,
close to $BD$, we define
	\begin{eqnarray}
	  U= e^{\kappa_{1} u} \; ,
	\label{cmc-e190}
	  \\
	  V= - e^{- \kappa_{1} v}
	\label{cmc-e200}
	\end{eqnarray}
so that $V$ tends to zero as $v$ tends to plus infinity. 
Near to the (future) cosmological event horizon the metric in these
coordinates becomes
	\begin{equation}
	  ds^{2} = \frac{2 r_{1}}{\kappa_{1}} dU dV
	  + r^{2} d \Omega^{2} \; ,
	\label{cmc-e210}
	\end{equation}
which is regular at the cosmological event horizon.

%
%
\begin{figure}[h]
\leavevmode
\begin{center}
\epsfxsize=0.3\textwidth
\leavevmode\epsffile{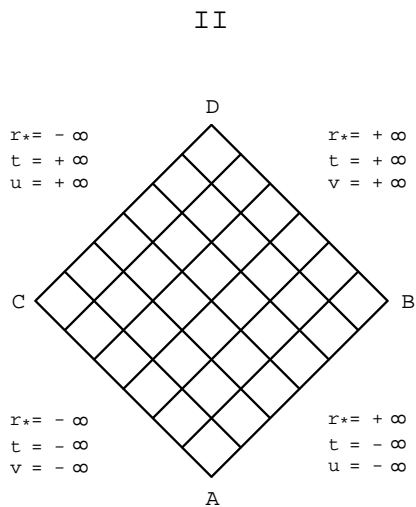}
\end{center}
\vspace{0.5cm}
\caption{
Part of the conformal diagram of {\protect Fig.~\ref{cmc-f10}}
showing region II and details of its causal structure (see the
text for details).}
\label{cmc-f30}
\end{figure}

\subsection*{Region III}

Figure~\ref{cmc-f40} depicts region III of Fig.~\ref{cmc-f10}.
We define two radial null coordinates $(u,v)$ as before,
defined by Eqs.(\ref{cmc-e130}) and (\ref{cmc-e140}). Surfaces
of constant $u$ and $v$ in Fig.~\ref{cmc-f40} are shown 
as lines at $45^{o}$ to the horizontal, with $v$ running 
parallel to the Cauchy horizon $DF$ and $u$ running parallel
to the black hole event horizon $CD$. It should be noted 
that the way we have defined $u$ and $v$ in region III is
neither unique nor conventional~\cite{cmc-b80}. Indeed, whilst
$r_{*}$ is constrained [via Eq.~(\ref{cmc-e40})]
to take the values minus infinity along $DCE$ and plus
infinity along $DFE$, there is no such constraint on the
parameter $t$. With $u$ and $v$ as defined, $t$ assumes 
the values minus infinity along $CEF$ and plus infinity
along $CDF$, in order to preserve the requirement that
both the Cauchy horizon and the cosmological event horizon
be located at $v = + \infty$. One can equally well choose
the more conventional definitions, $u = r_{*} + t$ and
$v = r_{*} - t$, and have $t$ be minus infinity on $CDF$
and plus infinity on $CEF$~\cite{cmc-b90}. No convention
is strictly correct, but should be decided upon according
to the problem at hand. The convention we choose is
particularly well suited to the matching of waves across the event
horizon, travelling from region II to region III.
In looking at regular coordinate systems,
near the boundaries of region III, we shall restrict our
attention to $DF$, since the regular coordinates here
play a crucial role when examining fluxes of energy crossing
the Cauchy horizon, as we shall see in later sections. Near $DF$
the regular coordinates $(U,V)$ are defined by
	\begin{eqnarray}
	  U &=& e^{\kappa_{3} u} \; ,
	\label{cmc-e220}
	  \\
	  V &=& - e^{- \kappa_{3} v}
	\label{cmc-e230}
	\end{eqnarray}
so that $V$ tends to zero as $v$ tends to plus infinity. The 
metric near $DF$, then becomes
	\begin{equation}
	  ds^{2} = - \frac{2 r_{3}}{\kappa_{3}} dU dV
	  + r^{2} d \Omega^{2} \; ,
	\label{cmc-e240}
	\end{equation}
which is obviously regular at the Cauchy horizon.
%
%
\begin{figure}[h]
\leavevmode
\begin{center}
\epsfxsize=0.3\textwidth
\leavevmode\epsffile{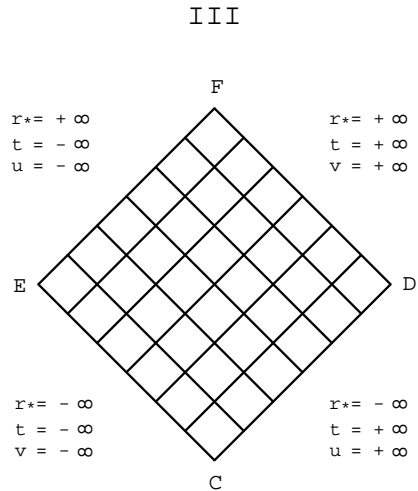}
\end{center}
\vspace{0.5cm}
\caption{
Part of the conformal diagram of {\protect Fig.~\ref{cmc-f10}}
showing region III and details of its causal structure (see the
text for details).}
\label{cmc-f40}
\end{figure}

\subsection{Radial Geodesics}

\label{cmc-s50}

The role of geodesic observers, particularly what they measure,
plays an important role in the study of any spacetime. 
In this section we shall consider certain concepts of 
geodesic observers that will be relevant to later sections.
For the sake of both clarity and conciseness, we restrict our attention
to radially free-falling observers $(d \theta = d \varphi = 0)$.
The first integrals of geodesic motion are
	\begin{eqnarray}
 	  E^{2} &=& \dot{r}^{2} + f \; ,
	\label{cmc-e250}
	  \\
	  \dot{v} &=& \frac{E \pm \sqrt{(E^{2} - f)}}{f}
	\label{cmc-e260}
	\end{eqnarray}
where E is the constant of motion (`energy') associated
with the timelike killing vector $\xi = \partial_{v}$,
which is given by
	\begin{equation}
	  E = - g_{\mu \nu} \xi^{\mu} u^{\nu} \; ,
	\label{cmc-e270}
	\end{equation}
and ${\bf u} = d / d \tau$ is the observer's four velocity.
The choice of sign in Eq.~(\ref{cmc-e260}) depends upon 
whether or not the observer is travelling on a path of
decreasing $r$, with respect to $\tau$, in which case
the negative root is taken, or whether they are on
a path of increasing $r$ in which case the positive root
is assumed. The path AA$'$ in Fig.~\ref{cmc-f10}
represents the typical world line of such an observer.
In region IV$'$ the observer, having been on a path
of decreasing $r$, reaches a turning point in her
(or his) motion and proceeds to move along a path
of increasing $r$. Eventually the observer reaches
region II$'$, another asymptotically de Sitter
universe isometric to region II. The occurrence of
a turning point in the evolution for 
$0 < r < r_{3}$ can be understood by examining the
radial geodesic equation~(\ref{cmc-e250}). If a turning
point exists then $\dot{r} = 0$. The problem is then to
ascertain at what, if any, value of $r > 0$ does this
occur. Setting $\dot{r} = 0$ in Eq.~(\ref{cmc-e250})
yields
	\begin{equation}
	  f(r) - E^{2} = 0 \; ,
	\label{cmc-e280}
	\end{equation}
the roots of which then define the turning points of
the evolution. Figure~\ref{cmc-f50} shows a plot of
$f(r)$ vs.~$r$ for the generic case of three distinct
spacetime horizons. The second plot is Eq.~(\ref{cmc-e280})
vs.~$r$, obtained by a vertical translation of $f(r)$ by
an amount $- E^{2}$. For realistic observers there is
one turning point in region IV$'$. Thus, there do exist 
paths for which radially infalling observers can
enter the black hole, pass by the timelike spacetime
singularity, and emerge in to another asymptotically
de Sitter universe.
%
%
\begin{figure}[h]
\leavevmode
\begin{center}
\epsfxsize=0.3\textwidth
\leavevmode\epsffile{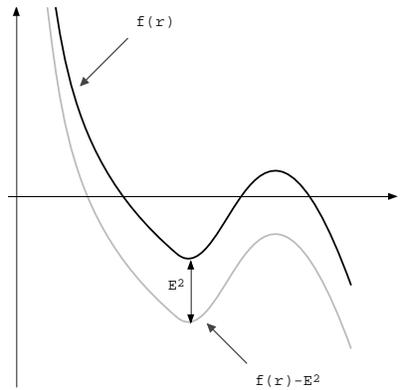}
\end{center}
\vspace{0.5cm}
\caption{
A graph of $f(r)$ vs.~$r$ and $f(r)-E^{2}$ vs.~$r$.
The three zeros in $f(r)$ (from right to left)
correspond to the spacetime
horizons $r_{1}, r_{2}$ and $r_{3}$. 
The zero in $f(r) - E^{2}$ at $0 < r < r_{3}$
corresponds to a turning point in the radial motion
of a freely falling observer inside region IV$'$, as
shown in Fig.~\ref{cmc-f10}.}
\label{cmc-f50}
\end{figure} 

We now turn our attention to observers that are
in the vicinity of either the cosmological
event horizon or the Cauchy horizon. The
importance of these observers, to the study of Cauchy
horizon stability, will
become transparent in later sections.
For clarity in what comes later we use subscripts to denote
which region we are considering.
For radially free-falling observers approaching 
the cosmological event horizon, it is easy to
show, using Eqs.~(\ref{cmc-e250}) and~(\ref{cmc-e260}),
that
	\begin{eqnarray}
	  \dot{r}_{\scriptscriptstyle II} &\simeq& 
	  | E_{\scriptscriptstyle II} | \; ,
        \label{cmc-e281} 
	  \\
	  \dot{v}_{\scriptscriptstyle II} &\simeq& 
	  \frac{2 |E_{\scriptscriptstyle II}|}{
	  f_{\scriptscriptstyle II}} \; ,
	\label{cmc-e282}
        \end{eqnarray}
The equation for $\dot{r}$ merely tells us that our
observer is approaching the future cosmological event 
horizon (increasing $r$) at a rate that depends upon
his initial energy. The  equation for $\dot{v}$ is
more interesting. As we approach the cosmological
event horizon $\dot{v}$ diverges, since $f$ tends to zero there. 
From  Eq.~(\ref{cmc-e110}), with $r = r_{1}$, we
can see that~\cite{cmc-b95}
	\begin{equation}
	 \dot{v}_{\scriptscriptstyle II} \equiv 
	 \frac{d v_{\scriptscriptstyle II}}{d 
	 \tau_{\scriptscriptstyle II}}
	 \simeq \frac{| E_{\scriptscriptstyle II} |}{ 
	 r_{1} \kappa_{1}}
	 e^{\kappa_{1} v} \ \ \ \ {\rm as} \ \ \ \ 
	 v \rightarrow \infty \; .
	\label{cmc-e283}
	\end{equation}
For observers near the Cauchy horizon, it can similarly
be shown that
	\begin{eqnarray}
	  \dot{r}_{\scriptscriptstyle III} & \simeq & 
	  - | E_{\scriptscriptstyle III} | \; ,
	\label{cmc-e284}
	  \\
	  \dot{v}_{\scriptscriptstyle III} & \simeq &  
	  -\frac{2 |E_{\scriptscriptstyle III}|}{
	  f_{\scriptscriptstyle III}} \; .
	\label{cmc-e285}
	\end{eqnarray}
Again we can see that $\dot{v}$ diverges as the
observer approaches the Cauchy horizon, and one can show that
	\begin{equation}
	  \dot{v}_{\scriptscriptstyle III} \equiv 
	  \frac{d v_{\scriptscriptstyle III}}{
	  d \tau_{\scriptscriptstyle III}}
	  \simeq \frac{|E_{\scriptscriptstyle III}|}{r_{3} \kappa_{3}}
	  e^{\kappa_{3} v} \ \ \ \ {\rm as} \ \ \ \ 
	  v \rightarrow \infty \; .
	\label{cmc-e286}
	\end{equation}
Since $\dot{r}$ approaches a fixed constant value as we approach
either horizon, the velocity of the observer will be dominated 
by the $u^{v}$ component. Of course the divergence
of $\dot{v}$ is due purely to the choice of coordinates, and
one can choose coordinates, such as the Kruskal-Szekeres
coordinates of Sec~\ref{cmc-s30}, that are regular at  
a particular horizon and in which the four velocity is
well behaved there.
%
%

\section{An Historical Overview}
\label{cmc-s60}

%
%

\subsection{Mellor and Moss 1990}
\label{cmc-s70}

The first investigative study of
Cauchy horizon stability in black hole-de Sitter spacetimes
was performed by Felicity Mellor and Ian Moss in
1990~\cite{cmc-b100}. Confining their attention to
the spherically symmetric Reissner-Nordstr\"{o}m-de
Sitter spacetime, Mellor and Moss considered the
effect of gravitational perturbations, generated
in the exterior of the black hole, on the Cauchy
horizon in the interior. They examined the flux 
of radiation due to these perturbations, as seen
by observers crossing the Cauchy horizon, and
concluded that for the cases they studied, the
Cauchy horizon is in fact stable to such
perturbations~\cite{cmc-b110}. 

Unfortunately a detailed exposition of the work
of Mellor and Moss would lead to quite a
{\it tour de force} in algebra and cause us to
stray from picture we are trying to paint here.
We will, therefore, sketch only an outline of
their investigation.

\subsection*{Analysis}

The method employed by Mellor and Moss is 
analogous to that used by Chandrasekhar 
and Hartle in their study of the Cauchy
horizon instability in the Reissner-Nordstr\"{o}m
spacetime~\cite{cmc-b120,cmc-b130}. The technique
is that of linear perturbations about the
background spacetime
	\begin{equation}
	  \tilde{g}_{\mu \nu} = g_{\mu \nu}
	  + \epsilon h_{\mu \nu} \; ,
	\label{cmc-e290}
	\end{equation}
where $\tilde{g}_{\mu \nu}$ represents the
components of the perturbed metric, $g_{\mu \nu}$ 
the components of the background metric and
$h_{\mu \nu}$ is the perturbation about the
background. The dimensionless quantity
$\epsilon$ is just an expansion parameter.
By constraining the perturbed metric to
be a solution to the Einstein-Maxwell 
equations (since there is a non-zero
background electromagnetic field) to
linear order in $\epsilon$, one obtains 
linearized equations for the metric perturbations
$h_{\mu \nu}$. These perturbations fall in to
two distinct classes
        \begin{description}
	  \item[Axial Perturbations : ]
	     Change sign under a change of sign of the 
	     azimuthal angle $\varphi$
	  \item[Polar Perturbations : ]
	     Invariant under a change of sign of $\varphi$
	\end{description}
With this definition it is easy to conclude, for
a spherically symmetric background spacetime, that the
polar perturbations have non-vanishing background values,
whereas the axial perturbations vanish on the
background. By considering the effect of the perturbations
on the Ricci tensor, along with the linearized form of
the Maxwell equations, it is quite remarkable that both
the axial and polar perturbations can be cast in to the 
form of a set of one-dimensional Schr\"{o}dinger type
wave equations, four in all,
	\begin{equation}
	  \frac{d^{2} Z_{j}^{\pm}}{d r_{*}^{2}}
	  + ( \sigma^{2} -V_{j}^{\pm} ) Z^{\pm}_{j} = 0
	  \ \ \  {\rm for} \ \ \ j=1,2 \; .
	\label{cmc-e300}
	\end{equation}
Here the $Z_{j}^{-}$ denote the two fields that describe 
the axial perturbations, $Z_{j}^{+}$ the polar perturbations
and $\sigma$ is the frequency of the perturbation modes
with time dependence $e^{i \sigma t}$. The functions 
$V_{j}^{\pm} ( r ; \ell, M, Q, \Lambda )$ are the
associated potentials, often referred to as the 
{\it effective potentials}, whose precise forms are too 
complicated to reproduce here but have been given by
Mellor and Moss~\cite{cmc-b100}. These equations are
generalizations of the Regge-Wheeler and Zerilli equations,
obtained from perturbations about asymptotically flat 
spacetimes~\cite{cmc-b140}.  One can, therefore, reduce the problem of
linear perturbations to an ensemble of one dimensional
scattering problems. In order to determine the whether
or not the Cauchy horizon is stable, one examines the
flux of radiation, due to the perturbations, encountered
by an observer crossing the Cauchy horizon. The behavior 
of the flux near the Cauchy horizon is ultimately dictated
by the scattering the fields have undergone in propagating
from the exterior to the interior, which is in turn
governed by the transmission $(T)$ and reflection $(R)$
coefficients. Close to the Cauchy horizon, the modes
$Z \in \{ Z^{\pm}_{j} \}$ have the asymptotic form
	\begin{equation}
	  Z \rightarrow A ( \sigma ) e^{-i \sigma r_{*}}
	  + B ( \sigma ) e^{i \sigma r_{*}} 
	  \ \ \ {\rm as} \ \ \
	  r \rightarrow r_{3} \; .
	\label{cmc-e310}
	\end{equation}
As an observer approaches the Cauchy horizon, their four-velocity
${\bf u}$ is proportional to $\partial_{V}$, where $V$ is the
Kruskal-Szekeres coordinate regular at the  Cauchy horizon [see
Sec.~\ref{cmc-s40}, Eq.~(\ref{cmc-e230})], and the flux of 
radiation he or she sees is proportional to
	\begin{equation}
	  \partial_{V} z(t) = \frac{1}{\kappa_{3}} 
	  e^{\kappa_{3} v} \partial_{v} z(t) \; ,
	\label{cmc-e320}
	\end{equation}
where $z(t)$ is the Fourier transform of the product of the
mode function $Z(\sigma,t)$ and the initial data
function $W(\sigma)$, namely
	\begin{equation}
	  z(t) = \frac{1}{2 \pi} \int W(\sigma) Z(\sigma,t)
	  e^{i \sigma t} d\sigma \; .
	\label{cmc-e330}
	\end{equation}
With the asymptotic form for the mode function, Eq.~(\ref{cmc-e310}),
the flux near the Cauchy horizon is proportional to
	\begin{equation}
	  \partial_{V} z(t) = -
	  \frac{i e^{\kappa_{3} v}}{2 \pi \kappa_{3}}
	  \int \sigma W(\sigma) A(\sigma) e^{-i \sigma v}
	  d\sigma \; .
	\label{cmc-e340}
	\end{equation}
It should be noted that the more conventional
definition for the null coordinates, $u$ and $v$, described in
Sec.~\ref{cmc-s40} have been used in obtaining Eq.~(\ref{cmc-e340}).
By implementing known results from one-dimensional
scattering theory~\cite{cmc-b120,cmc-b140,cmc-b150}, and matching 
modes, scattered across the black hole event horizon from region II
to region III, Mellor and Moss are able to show that
	\begin{equation}
	  \overleftarrow{A}_{\scriptsize III} (\sigma)
	  = \frac{\overleftarrow{T}_{\scriptsize II} (- \sigma )}{
	  \overleftarrow{T}_{\scriptsize III} ( -\sigma )}
	  \ \ \ {\rm and} \ \ \
	  \overrightarrow{A}_{\scriptsize III} (\sigma) 
	  = \frac{\overrightarrow{R}_{\scriptsize II} (-\sigma )}{
	  \overleftarrow{T}_{\scriptsize III} (-\sigma)} \; ,
	\label{cmc-e350}
	\end{equation}
where $\overleftarrow{Z}$ denotes an ingoing mode, originating 
from the (past) cosmological event horizon (AB in 
Fig.~\ref{cmc-f30}) in region II and $\overrightarrow{Z}$ 
denotes an outgoing mode, originating from the past black hole
event horizon (AC in Fig.~\ref{cmc-f30}). The subscripts,
of course, refer to the exterior and interior regions as
defined in Sec.~\ref{cmc-s40}, and differ from those
definitions in~\cite{cmc-b100}. The analytic structure of the
reciprocal transmission coefficients $T^{-1}_{\scriptsize II}$
and $T^{-1}_{\scriptsize III}$ are obtainable directly from the
work of Chandrasekhar and Hartle~\cite{cmc-b120}. Both
$T^{-1}_{\scriptsize II}$ and $T^{-1}_{\scriptsize III}$ have poles
along the imaginary $\sigma$ axis at integer multiples
of $i \kappa_{+}$ and $i \kappa_{-}$, where $\kappa_{-}$
and $\kappa_{+}$ are the surface gravities (Sec.~\ref{cmc-s30})
of the horizons on the incident and transmission
sides of the potential respectively. Due to zeros in 
$T_{\scriptsize II}$ and $R_{\scriptsize II}$ cancelling the pole
in $T_{\scriptsize III}$, the leading pole in $A(\sigma)$  comes
from a pole at $i \kappa_{3}$, whose residue results in a flux at the
Cauchy horizon which is finite. The only other poles that could
enter $A(\sigma)$ would come from $T_{\scriptsize II}$ or 
$R_{\scriptsize II}$.  Using numerical techniques, Mellor and Moss provided
convincing evidence that there are no poles in either 
$T_{\scriptsize II}$ or $R_{\scriptsize II}$ in the range 
$ 0 < {\rm Im}(\sigma) < \kappa_{3} $, thus concluding that the
Cauchy horizon in a Reissner-Nordstr\"{o}m-de Sitter black hole
is stable to linear perturbations.

\subsection*{Remarks}

There is one very subtle assumption 
in this analysis, which was only recognized later in the work of
Brady and Poisson~\cite{cmc-b160}. In studying 
the Reissner-Nordstr\"{o}m-de Sitter interior, Mellor and Moss
had not considered what constituted reasonable initial data
in the exterior. We shall see that this is tantamount to 
ignoring the pole structure of the Fourier transform of the 
initial amplitude of the field modes $W(\sigma)$. By doing this
there is an implicit assumption of vanishing flux at the
cosmological event horizon which, on physical grounds, seems
to be a rather strong restriction on the initial data. The
more reasonable requirement of a finite, but non-zero, flux
of energy at the cosmological event horizon still confirms 
that the Cauchy horizon can be stable, but only for a
restricted  family of solutions as we shall see next.

%
%

\subsection{Brady and Poisson 1992}
\label{cmc-s80}

The question of Cauchy horizon stability in the
Reissner-Nordstr\"{o}m-de Sitter spacetime was
revisited in 1992 by Patrick Brady and Eric
Poisson~\cite{cmc-b160}. Rather than contend
with the formidable algebra of the gravitational
perturbation method employed by Mellor and Moss,
Brady and Poisson pursued a much
simpler approach to the problem. Their model, a
generalization of the study of perturbed
a Reissner-Nordstr\"{o}m interior due to 
Hiscock~\cite{cmc-b170}, mimics the perturbations 
propagating from region II to region III, across
the event horizon, as a spherical inflow of
null dust. Whilst an exact solution to the
Einstein equations (in the presence of a
cosmological constant) exists~\cite{cmc-b180},
Brady and Poisson instead treat the inflow
as a linear perturbation, with the null dust
propagating on a fixed Reissner-Nordstr\"{o}m-de 
Sitter background. The beauty of the model is
its simplicity together with its ability
to capture the essential features of the Cauchy
horizon stability issue. Whilst the model does 
not lend itself to a detailed investigation of the
interior, it does incorporate the subtlety of the
initial data problem missed in the Mellor-Moss
analysis and allows us to make the following
three, important,  conclusions
	\begin{itemize}
	  \item
	  The infinite time compression effect,
	  discussed in the introduction, is a 
	  {\it sufficient} requirement for an
	  instability of the Cauchy horizon, but it
	  is not a necessary requirement. A
	  {\it necessary} condition is an infinite
	  compression of the ratio of differential
	  proper times.
	  \item
	  The requirement of a finite
          but non-zero flux of energy crossing
	  the cosmological event horizon (initial data) leads
	  to a significantly different stability
	  condition than that obtained by Mellor 
	  and Moss. Namely the Cauchy horizon
	  is stable provided that the surface 
	  gravity there is less than that at the
	  cosmological event horizon,
		\begin{equation}
		  \kappa_{1} > \kappa_{3} \; .
                \label{cmc-e360}
		\end{equation}
	  \item
	  If a backreaction calculation similar to
	  that used by Poisson and Israel~\cite{cmc-b190}
	  to study the interior of Reissner-Nordstr\"{o}m
	  were performed, the model suggests the following
	  conclusions
	    \begin{enumerate}
	      \item
	       If $\kappa_{3} > 2 \kappa_{1}$ then there 
	       would be a mass inflation type singularity.
	      \item
	       If $2 \kappa_{1} > \kappa_{3} > \kappa_{1}$
	       then there would be  a divergent flux at
	       the Cauchy horizon but the internal mass
	       function would approach some finite 
	       asymptotic value.
	     \end{enumerate}
	\end{itemize}

\subsection*{Analysis}

The inflow of spherical null dust propagating on the
fixed background, described by Eqs.~(\ref{cmc-e10}) and
(\ref{cmc-e20}),  is characterized by the stress-energy
tensor
	\begin{equation}
	  T_{\mu \nu} =  \frac{L(v)}{4 \pi r^{2}}
	  l_{\mu} l_{\nu} \; ,
	\label{cmc-e380}
	\end{equation}
where $L(v)$ is the {\it luminosity} function and $l_{\mu} =
- \partial_{\mu} v$ is a vector tangent to ingoing radial
null geodesics. An observer crossing the inflow, with a four 
velocity $\bf u$, measures a flux of energy $\rho$ given by
	\begin{equation}
	  \rho \equiv T_{\mu \nu} u^{\mu} u^{\nu} \; .
	\label{cmc-e390}
	\end{equation}
A radially free-falling observer in the vicinity of the
cosmological event horizon measures, therefore, a flux
	\begin{equation}
	  \rho_{\scriptscriptstyle II} \simeq T_{vv} \dot{v}^{2} 
	  = \frac{| E_{\rm \scriptscriptstyle II} |^{2}}{4 \pi r_{1}^{2}}
	  L(v) e^{2 \kappa_{1} v} \; ,
	\label{cmc-e400}
	\end{equation}
where we have used Eq.~(\ref{cmc-e283}) of Sec.~\ref{cmc-s40}. On physical
grounds, we expect the flux of energy at the cosmological
event horizon to be finite and in general non-zero. This requires
that
	\begin{equation}
	  L(v) = K(v) e^{-2 \kappa_{1} v} \; ,
	\label{cmc-e410}
	\end{equation}
such that
	\begin{equation}
	  \lim_{v \rightarrow \infty} K(v)
	  \equiv K_{\infty} \neq 0 \; .
	\label{cmc-e420}
	\end{equation}
Brady and Poisson refer to this later condition as the 
{\it minimal requirement}, which ensures a non-zero
flux is measured by the observer at the cosmological 
event horizon.

A radially infalling observer in the proximity of the Cauchy
horizon measures a flux of energy $\rho_{\scriptscriptstyle III}$,
due to the inflow, given by
	\begin{eqnarray}
	  \rho_{\scriptscriptstyle III} &=&
	  \frac{| E_{\scriptscriptstyle III} |^{2}}{4 \pi r_{3}^{2}}
	  L(v) e^{2 \kappa_{3} v} \nonumber \\
	  &=& 
	  \frac{| E_{\scriptscriptstyle III} |^{2}}{4 \pi r_{3}^{2}}
	  K(v) e^{2 ( \kappa_{3} - \kappa_{1} ) v} \; ,
	\label{cmc-e430}
	\end{eqnarray}
using Eq.~(\ref{cmc-e410}). Thus, at the Cauchy horizon,
if $\kappa_{3} > \kappa_{1}$ the energy
density diverges, indicating an instability of the 
horizon~\cite{cmc-b110}. On the otherhand, if 
$\kappa_{1} > \kappa_{3}$ then the energy density is
finite, suggesting the horizon is stable.

\subsection*{Remarks}
%
%
\begin{figure}[h]
\leavevmode
\begin{center}
\epsfxsize=0.4\textwidth
\leavevmode\epsffile{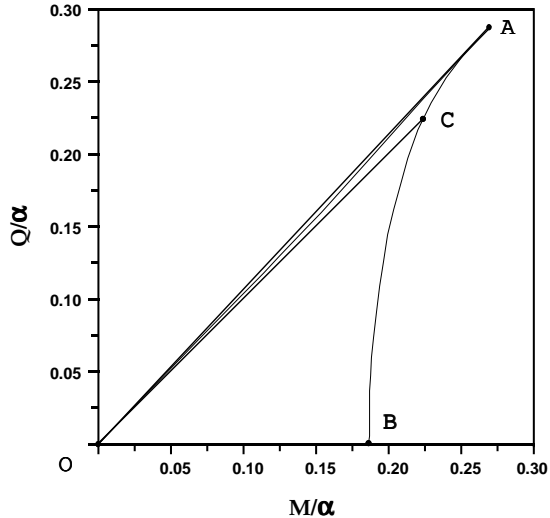}
\end{center}
\caption{A plot of the black hole parameter space showing the
region of stability defined by {\protect Eq.~(\ref{cmc-e360})}.
The upper line OA denotes the condition $r_{2} = r_{3}$ whilst
the line AB denotes the condition $r_{1} = r_{2}$. The lower
line OA, just visible, denotes the condition $\kappa_{1} =
\kappa_{3}$. The narrow region between the lines OA is
the valley of stability. The line OC denotes the condition
$|Q|=M$ showing that the stability condition requires
$Q>M$ which can always be achieved by adding charge to the black hole.}
\label{cmc-f60}
\end{figure}

The stability condition of Eq.~(\ref{cmc-e360}) offered by the
Brady-Poisson model is more restrictive than the original
proposal of Mellor and Moss, and one might ask whether or
not this condition can be met in any realistic sense. 
Figure~\ref{cmc-f60} shows the region in the parameter
space $( M, Q, \Lambda)$ for which the stability condition
holds. Though small, the {\it valley of stability} is
of finite, non-zero, measure on the parameter space.
An important point relating to this plot is  that 
even if $\Lambda$ is small, provided it is non-zero,
one can always find solutions with a stable Cauchy horizon.
There is absolutely no restriction to having an 
unacceptably large value of $\Lambda$~\cite{cmc-b200}.
Another point worth clarifying is that the stability condition
requires $Q^{2} > M^{2}$. Unlike the Reissner-Nordstr\"{o}m
black hole, there is no physical reason why one cannot
keep putting charge on a Reissner-Nordstr\"{o}m-de Sitter
black hole to the point where the charge exceeds the 
mass~\cite{cmc-b210}. In {\it practice}
one could fine tune the black hole, by dropping in charge
and/or mass, until it achieved a state lying in the 
stability region, allowing safe passage across the
Cauchy horizon. 

The stability condition also has the endearing quality
that it can be understood intuitively, in terms of competing
redshift and blueshift effects. At the cosmological
event horizon there is an infinite redshift effect due to the
cosmological expansion, whilst at the Cauchy horizon, due
to its causal nature, there is an infinite blueshift effect. 
If the redshift outweighs the blueshift $(\kappa_{1} > \kappa_{3})$
then the measured energy density at the Cauchy horizon is
sufficiently diluted to render the horizon stable. 
Conversely, if the blueshift prevails over the redshift
$(\kappa_{3} > \kappa_{1})$ then the measured energy density
is sufficiently dense to render the horizon unstable.
Up until this juncture, the stability (or instability) of the Cauchy 
horizon had always been explained in terms of the time compression
effects between the exterior and interior, as discussed in
the Introduction. The Brady-Poisson
analysis demonstrates that while we can consider this a sufficient 
condition, it is not a necessary condition. For $\kappa_{1} <
\kappa_{3}$, the Cauchy horizon is unstable, but there is 
no infinite proper-time compression since the black hole 
is part of a de Sitter, i.e., closed, universe. What is necessary
is an infinite compression of the ratio of {\it differential}
proper-times, since blueshift and redshift are precisely
differential proper-time effects (See Appendix~\ref{cmc-a10}.)

The existence of a region in the 
parameter space for which the Cauchy horizon is unstable, but 
there is no associated growth in the internal mass function,
lead Brady and Poisson to speculate that if backreaction 
were taken in to account~\cite{cmc-b190}, then the Cauchy
horizon singularity might remain weak, in the sense that
no scalars formed from the curvature tensors diverge there.
Indeed, the Weyl scalar $\Psi_{2}$, that in spherical
symmetry is characterized by the mass function, is regular at 
the Cauchy horizon. However, we shall in Sec.~\ref{cmc-s100}
that there is in fact a scalar curvature singularity at the 
horizon, characterized by the divergence of the Kretchsmann
invariant $R_{\mu \nu \gamma \sigma} R^{\mu \nu \gamma \sigma}$.

%
%

\subsection{Mellor and Moss 1992}
\label{cmc-s90}

In light of the work by Brady and Poisson, Mellor 
and Moss have reassessed their stability analysis~\cite{cmc-b220}.
By requiring that observers crossing
the cosmological event horizon should measure
a finite but non-vanishing flux of energy,
it is relatively easy to show that the gravitational
perturbation approach to the Cauchy horizon problem
reproduces the stability condition, Eq.~(\ref{cmc-e360}),
in complete agreement with the Brady-Poisson
model.

\subsection*{Analysis}

The initial data requirement of a finite but non-vanishing
flux of energy being measured at the Cauchy horizon is
easily incorporated in to the gravitational perturbation
method by imposing pole structure on the
initial data function $W(\sigma)$, defined in Eq.~(\ref{cmc-e330}). 
Close to the cosmological event horizon, the modes have
the asymptotic form
	\[
	  Z \rightarrow e^{i \sigma r_{*}}
	  + R(\sigma)  e^{- i \sigma r_{*}} \; .
	\]
The flux of energy [Eq.~(\ref{cmc-e320})] near the
cosmological event horizon is thus proportional to
	\begin{equation}
	   \partial_{V} z(t) = 
           \frac{i e^{\kappa_{1} v}}{2 \pi \kappa_{1}}
           \int \sigma W(\sigma) e^{i \sigma v}
	   d\sigma \; ,
	\label{cmc-e440}
	\end{equation}
Now, if $W(\sigma)$ has any poles in the range 
$0 < \sigma < i \kappa_{1}$, then the flux would 
diverge, indicating an unphysical instability of the
cosmological event horizon. On the otherhand,
if there are no poles in the range $0 < \sigma \leq
i \kappa_{1}$  the flux is vanishing at the
cosmological horizon. From a physical point of
view this is a very strong requirement on
the initial data. If we impose the more reasonable
requirement of a finite, non-vanishing flux at
the cosmological event horizon then $W(\sigma)$
is forced to have a pole at $\sigma = i \kappa_{1}$.
Using the results of Sec.~\ref{cmc-s70}, this implies
that the Cauchy horizon will be stable provided that
$\kappa_{1} > \kappa_{3}$, as initially suggested by Brady and
Poisson.

\subsection*{Remarks}

From the analysis above it is clear that 
neglecting the  relevant pole structure of the
initial data, encoded in $W(\sigma)$, imposes
an unreasonably strong restriction on the 
initial perturbations, namely the flux
[Eq.~(\ref{cmc-e440})] at
the cosmological event horizon, due
to these perturbations, vanish. In relation
to the Brady-Poisson model, this condition
would require 
	\[
	  \lim_{v \rightarrow \infty}
	  K(v) = 0 \; ,
        \]
which can be seen, via Eq.~(\ref{cmc-e430}), to 
lead to stability in all cases and clearly
explains the original result of Mellor and Moss.

%
%

\subsection{Brady, N\'{u}\~{n}ez and Sinha 1993}
\label{cmc-s100}

Unlike the pure inflow model of Brady and Poisson, 
the perturbative
content of the analysis by Mellor and Moss does
incorporate the effects of field-scattering off the spacetime
curvature. However, neither analysis takes account of 
non-linear backreaction effects produced by
the field evolution. The first, and so far only, 
calculation to consider the effects of 
backreaction on the geometry was provided by
Brady, N\'{u}\~{n}ez and Sinha~\cite{cmc-b230} in 1993.

Generalizing the backreaction model devised by
Poisson and Israel~\cite{cmc-b190}, for their
study of the Reissner-Nordstr\"{o}m interior,
Brady {\it et al.} arrived at the following
important conclusions
	\begin{itemize}
	  \item
          The Cauchy horizon in Reissner-Nordstr\"{o}m-de Sitter
	  is stable provided that
	    \[
	      \kappa_{1} > \kappa_{3} \; .
	    \]
          \item
	  For $2 \kappa_{1} > \kappa_{3} > \kappa_{1}$
	    \begin{enumerate}
	      \item
	      There is a divergent flux of energy at the 
	      Cauchy horizon but no corresponding 
	      growth of the internal mass function.  
	      \item
	      The curvature singularity at the Cauchy 
	      horizon is {\it strong}, revealing itself 
	      as a divergence of the Kretchsmann scalar 
	      $R_{\mu \nu \gamma \sigma} R^{\mu \nu \gamma \sigma}$.
            \end{enumerate}
	  \item
	  For $\kappa_{3} > 2 \kappa_{1}$ the Cauchy horizon
	  instability is qualitatively similar to that found
	  in the Reissner-Nordstr\"{o}m solution.
	\end{itemize}

\subsection*{Analysis}

The analysis of Brady {\it et al.}, whilst simple, does require
a suitable knowledge of certain background material, which, for
contextual reason, we have not presented here. 
In fact, the generalization of the Poisson-Israel model to
include a cosmological constant is so straight forward
that much of what one would learn from it can be gained from
an understanding of the Poisson and Israel analysis~\cite{cmc-b240}. 
For that 
reason we shall not attempt to reproduce an outline of
the work by Brady {\it et al}. For the interested reader
Appendix~\ref{cmc-a20} details a simpler picture of the 
Brady-N\'{u}\~{n}ez-Sinha backreaction model, 
originally due to Ori~\cite{cmc-b250}, which captures some of
the essence of their analysis.

\subsection*{Remarks}

The relevance of the work by Brady, N\'{u}\~{n}ez and Sinha,
to the study of Cauchy horizon stability in 
Reissner-Nordstr\"{o}m-de Sitter spacetimes, cannot
be over emphasized. As the only model, to date, to
incorporate the effects of backreaction, it is both a
natural and important extension to the preceding
analyses. The agreement between the linear and non-linear
studies is pleasing and, at present, is the best evidence we have toward
the conjecture that the Cauchy horizon in black hole-de Sitter
spacetimes is stable for $\kappa_{1} > \kappa_{3}$. However,
we should be careful not to paint too much of a rosy picture
here. Two possible caveats to the work of Brady {\it et al.} that
have direct bearing on the conjecture are
	\begin{enumerate}
	  \item
	  While the model includes backreaction, it
	  does restrict these effects to a spherically
	  symmetric spacetime.
	  \item
	  The model imposes the minimal requirement, of
	  Brady and Poisson, on the fields.
	\end{enumerate}
The first caveat is extremely difficult to address. To be more clear
about this point, we have to ask what it is we are trying to prove
in our studies of the Cauchy horizon in black hole-de 
Sitter spacetimes. We can pose this as a question;
	\begin{quote}
          Can generic collapse, in de Sitter space, 
          lead to a black hole with a stable Cauchy horizon?
	\end{quote}
It is difficult, if not impossible, to answer this question. Currently
our best approach is to investigate more tractable models of the 
interior, such as those restricted to spherical symmetry, to guide
our understanding and hopefully one day answer our questions.
To what extent the backreaction
will play a role in the stability issue is unclear. Indeed, since the
curvature appears to be regular at the Cauchy horizon (in the
case of stability) we might expect that the backreaction 
plays a minor role in the shaping of the spacetime 
interior. However, it may
turn out that in a fully non-linear simulation of the collapse
that the backreaction becomes important at early times, 
causing the interior to be significantly different to what
our models predict.  While we cannot provide definite
answers to this and similar questions, we can gain some
insight from our current models. With the conclusions of
the linear model by Mellor and Moss in conjunction with
the results of Brady {\it et al.}  it is clear that
the backreaction is not playing a significant role in
the evolution of the spacetime close to the Cauchy horizon, 
at least within the confines of spherical 
symmetry and null dust flow. 
What about axisymmetric spacetimes? In 
Sec.~\ref{cmc-s110} we shall discuss the details of a recent
perturbation analysis of the
Kerr-de Sitter spacetime, similar to that implemented
by Mellor and Moss in their study of the Reissner-Nordstr\"{o}m-de Sitter
spacetime. The results of this analysis suggest that rotation has no
effect on the stability conjecture, within the perturbative 
approach employed.  Unfortunately no backreaction calculation, akin to that
of Brady {\it et al.}, currently exist for  Kerr-de Sitter spacetimes,
against which we could compare the linear analysis. We shall have
to wait to see whether backreaction, away from the confines of spherical,
in anyway alters our current picture of stability.

The second caveat is more readily addressed.  The 
requirement of a finite but non-vanishing flux
at the Cauchy horizon leads to Eqs.~(\ref{cmc-e410}) 
and (\ref{cmc-e420}) for the luminosity function $L(v)$.
There is little doubt that this is the
most physically reasonable condition one could impose on
the initial perturbations. For the case of black
holes in asymptotically flat black hole spacetimes
there is no such requirement on the luminosity
function~\cite{cmc-b260}. In fact, requiring the
perturbations to vanish at future null infinity $(v = r = \infty)$
just requires the luminosity to be a decreasing function
of advanced time. The exact form for the luminosity
function is arrived at from an analysis of
the late time behavior of fields in the exterior due to
Price~\cite{cmc-b270} and Bi\v{c}\'{a}k~\cite{cmc-b280}.
For black holes in de Sitter space, the minimal requirement is
implicitly imposing the form on the late time behavior of
fields in the vicinity of the cosmological
horizon. In Sec.~\ref{cmc-s150} we will discuss the
details of a numerical investigation in to the
late time behavior of fields in the Schwarzschild-de Sitter
and Reissner-Nordstr\"{o}m-de Sitter spacetimes and
comment there on its compatibility with the minimal requirement
of Brady and Poisson.

For $2 \kappa_{1} > \kappa_{3} > \kappa_{1}$, the non-linear
analysis shows the remarkable structure alluded to by
Brady and Poisson~\cite{cmc-b160} -- a divergent flux but
no corresponding mass inflation.
In this instance, the Weyl
scalar $\Psi_{2}$ no longer acts as a measure of the
divergence as it does in the asymptotically flat case, instead
it is the Kretschmann invariant that signals the existence of
a scalar (strong) singularity at the Cauchy horizon.
It had been assumed
that the existence of a divergent flux at the Cauchy horizon
would always be met by an associated growth in the internal
mass parameter. That mass-inflation does not occur in this
parameter range makes the black hole-de Sitter case all the more
interesting and worthy of study.

%
%

\subsection{Chambers and Moss 1994}
\label{cmc-s110}

Until 1994, all analyses of the Cauchy horizon in
black hole-de Sitter spacetimes had, for
simplicity, confined their
attention to the spherically symmetric
Reissner-Nordstr\"{o}m-de Sitter solution.
However, in general, one expects a black hole
formed in a realistic gravitational collapse 
situation to be rotating and uncharged.  
It is therefore,  both natural and physically
well motivated to consider how the inclusion of
rotation might affect the current spherical
picture of Cauchy horizon stability.
The generalization  of the stability analysis 
by Mellor and Moss to the case of a rotating but
uncharged black hole was performed in 1994 by Chris
Chambers and Ian Moss~\cite{cmc-b285}.
Studying linear perturbations of scalar,
electromagnetic and gravitational fields on the
Kerr-de Sitter spacetime, Chambers and Moss
were able to conclude that the Cauchy horizon
was stable, provided that $\kappa_{1} > \kappa_{3}$,
as in the spherical case.
The method employed by Chambers and Moss is identical
to that used by Mellor and Moss in their study of the
Reissner-Nordstr\"{o}m-de Sitter spacetime (Sec.~\ref{cmc-s70}).
The essence of the approach is to transform the equations
for the perturbations to a set of one dimensional
Schr\"{o}dinger-type wave equations, thus reducing the
problem of linear perturbations to the more familiar
problem of one dimensional scattering. The entire
analysis is, as in the Mellor and Moss study, a
{\it tour de force} in algebra. Consequently we
sketch only an outline of the analysis.

\subsection*{Analysis}

Chambers and Moss consider three types of field 
perturbation
	\begin{itemize}
	  \item
	  Scalar perturbations (spin = 0)
	  \item
	  Electromagnetic perturbations (spin = 1)
	  \item
	  Gravitational perturbations (spin = 2)
	\end{itemize}
The scalar perturbations are represented as
a massless, minimally coupled scalar field obeying
$\Box \phi =0$, propagating on a fixed background described 
by the Kerr-Newman-de Sitter spacetime. Using separable
solutions of the form
	\begin{equation}
	  \phi = R(r) S(\mu) e^{-i \omega t} e^{i m \varphi} \;
	\label{cmc-e445}
	\end{equation}
where $\mu =a \cos (\theta)$ and $a$ is the rotation parameter.
Whilst no explicit solution to the angular equation exist,
for $\Lambda =0$ the solutions $S(\mu)$ are spheroidal
wave functions. The equation for the radial function
$R(r)$ is easily reduced to the form of a one dimensional
scattering equation
	\begin{equation}
	  \frac{d^{2} Z}{d r_{*}^{2}} + V  Z = 0
	  \; ,
	\label{cmc-e450}
	\end{equation}
where $Z(r) = (r^2 + a^{2})^{1/2} R(r)$, $r_{*}$ is the
analogous coordinate to Eq.~(\ref{cmc-e40}) for Kerr-Newman-de Sitter
and $V$ is the effective potential.

The electromagnetic perturbations on a Kerr-de Sitter
background (no background electromagnetic field),
are governed by Maxwell's equations. Chambers and Moss
resorted to the Newman-Penrose (NP) formalism to
describe these equations~\cite{cmc-b290}. In the NP formalism,
the six independent components of the field strength
tensor $F_{\mu \nu}$, are described by the three complex
Maxwell scalars $(\phi_{0}, \phi_{1}, \phi_{2})$. Trying
separable solutions of the form Eq.~(\ref{cmc-e445}), 
for each of the fields, enables one to write the
equations for the radial functions as a set of one
dimensional scattering equations similar to Eq.~(\ref{cmc-e450}).
The details of this transformation are quite involved and
the can be found in~\cite{cmc-b310}.

The gravitational field in the NP formalism is described
by the five complex Weyl scalars $(\Psi_{0}, \Psi_{1}, \Psi_{2},
\Psi_{3}, \Psi_{4})$, representing the ten degrees of freedom of
the gravitational field. Due to the nature of the Kerr-de Sitter
spacetime~\cite{cmc-b320}, $\Psi_{2}$ is the only non-vanishing
Weyl scalar on the background and only the remaining Weyl
scalars can be used to describe the gravitational perturbations.
While $\Psi_{0}$ and $\Psi_{4}$ are gauge invariant, and hence 
measurable, $\Psi_{1}$ and $\Psi_{3}$ are not
and, with a judicious choice of gauge, can
be made to vanish. With a significant amount of 
algebra~\cite{cmc-b330}, the equations for $\Psi_{0}$
and $\Psi_{4}$ can, by assuming seperable solutions of the form
Eq.~(\ref{cmc-e445}), be cast in the form of Eq.~(\ref{cmc-e450}).
Again, for the details, one is guided to~\cite{cmc-b310}.

With the equations governing the perturbations transformed
to a set of one dimensional scattering equations, much of
the hard work is over. An investigation of the flux of
energy, due to the perturbations, near the Cauchy horizon
is performed in an identical way to that implemented by 
Mellor and Moss. Imposing the minimal requirement of
Brady and Poisson and examining the pole structure
of the transmission and reflection coefficients, one
can show~\cite{cmc-b285} that the Cauchy horizon in Kerr-de Sitter
is stable provided that, as in the spherical case, 
	\[
	  \kappa_{1} > \kappa_{3} \; ,
	\]

\subsection*{Remarks}

The analysis of Chambers and Moss is the first to indicate that
Cauchy horizon stability, in black hole-de Sitter spacetimes,
is not an artefact of spherical
symmetry. Indeed, this analysis provides the best evidence yet
that generic collapse in de Sitter space can yield  black
holes with stable Cauchy horizons, violating the spirit (if
not the letter) of the strong cosmic censorship hypothesis.
Unfortunately, there is currently no calculation that takes 
in to account the  effects of backreaction on the spacetime.
Thus we have no model against which we can test the
predictions of the linear analysis. However, there is
little evidence to suggest that a backreaction
calculation would lead us to any other conclusions.
The valley of stability in the Kerr-de Sitter case is similar
to the charged case, shown in Fig.~\ref{cmc-f60}. One peculiar
property of the Kerr-de Sitter spacetime is the existence
of closed timelike curves, near the ring singularity at $r=0$.
These arise due to the timelike nature of the azimuthal
angle $\varphi$ for small values of $r$. The possibility
of observers crossing in to this region (IV) faces us with
the problem of doing physics in the presence of closed timelike curves.
There are many interesting aspects of the Kerr-de Sitter 
spacetime which have not received attention but provide
thought provoking possibilities~\cite{cmc-b310}.

%
%

\subsection{Markovi\'{c} and Poisson 1995}
\label{cmc-s120}

While we are primarily concerned with the classical
stability of the Cauchy horizon in black hole-de
Sitter spacetimes, we would not be presenting a fair
picture if we did not include the results of a
quantum analysis.

Classically we have seen that the Cauchy horizon can
be stable for a certain region of the black hole parameter
space $(M,Q,J,\Lambda)$, which leads to problems 
associated with the loss of predictability that occurs
beyond the Cauchy horizon. 
It is natural, therefore, to ask 
whether or not the Cauchy horizon can be quantum
mechanically stable or whether quantum effects
will restore predictability. So, how do the fluxes of quantum
fields affect the spacetime in
the vicinity of the Cauchy horizon? The intriguing
answer to this question was provided by
Dragoljub Markovi\'{c} and Eric Poisson~\cite{cmc-b340}
in 1995. Examining the quantum fluxes measured by an
observer approaching the Cauchy horizon, Markovi\'{c}
and Poisson were able to conclude that the horizon
is quantum mechanically unstable, except for the 
set of zero measure solutions with
	\[
	 \kappa_{1} = \kappa_{3} \; ,
	\]
which are represented by the lower line $OA$ in the parameter
space plot of Fig.~\ref{cmc-f60}.
\subsection*{Analysis}

To investigate the quantum stability of the Cauchy
horizon in a black hole spacetime generally requires
a knowledge of $\langle T_{\mu \nu} \rangle$, the renormalized
expectation value of the stress-energy tensor associated
with the quantum field. However, even within the confines of
spherical symmetry,
four dimensional calculations of $\langle T_{\mu \nu} \rangle$ are
extremely difficult and, for the case of black hole-de Sitter
spacetimes, the situation is especially difficult because
one cannot choose from the standard vacuum states~\cite{cmc-b350}, 
such as the Hartle-Hawking or Unruh states. Markovi\'{c}
and Poisson instead consider a simpler, but still instructive,
approach to the problem, quantizing a conformally invariant scalar
field on a two dimensional version of the Reissner-Nordstr\"{o}m-de
Sitter spacetime
	\begin{equation}
	  ds^{2} = -f(r) du dv\; ,
	\label{cmc-e460}
	\end{equation}
where the pair $(u,v)$ are the null coordinates defined in 
Sec.~\ref{cmc-s40} and $f(r)$ is given by Eq.~(\ref{cmc-e20}).
A quantum state that is regular on
both the cosmological horizons and the black hole event
horizons, for the two-dimensional case, has been provided
by Markovi\'{c} and Unruh~\cite{cmc-b370}. For this state,
the renormalized expectation value of the scalar field can 
be written as~\cite{cmc-b380}
	\begin{equation}
	  \langle T_{\mu \nu} \rangle = \theta_{\mu \nu} + t_{\mu \nu}
	  + (48 \pi)^{-1} R g_{\mu \nu} \; ,
	\label{cmc-e470}
	\end{equation}
where $R$ is the Ricci scalar associated with the two-dimensional
metric [Eq.~(\ref{cmc-e460})], $\theta_{\mu \nu}$ is a state
independent object and $t_{\mu \nu}$ is a state dependent 
object. Examining the expectation value of the energy density
measured by a freely falling observer crossing the Cauchy horizon,
$\langle \rho \rangle = \langle T_{\mu \nu} \rangle
u^{\mu} u^{\nu}$, in a regular coordinate frame reveals that
	\begin{equation}
	  \langle \rho_{\scriptscriptstyle  III} \rangle
	  = \frac{| E |^{2}}{48 \pi} 
	  (\kappa_{1}^{2} - \kappa_{3}^{2})
	  e^{2 \kappa_{3} v} \; ,
	\label{cmc-e480}
	\end{equation}
as $v$ tends to infinity [cf. Eq.~(\ref{cmc-e430})]. Thus, unless 
$\kappa_{1} = \kappa_{3}$, the Cauchy horizon in
Reissner-Nordstr\"{o}m-de Sitter is quantum mechanically unstable.

\subsection*{Remarks}

While the calculation is restricted to a simpler 
two-dimensional model, Markovi\'{c} and Poisson give 
a physical interpretation of Eq.~(\ref{cmc-e480}), 
in terms of the thermal quanta emitted by the horizons
and the gravitational redshifts and blueshifts such quanta
undergo, that suggests the result should hold true in the
four dimensional case and irrespectively of any 
particular quantum field.

The quantum mechanical instability of the  Cauchy horizon is
interesting for many reasons, but in particular it
is that this situation demonstrates how, even in regions
of spacetime where classical curvatures are not
necessarily large, quantum effects can be important. One
normally expects quantum effects to become important only when 
curvatures are sufficiently high (approaching planck scales). 
This result is similar to the situation in which a spacetime 
possesses regions containing closed timelike curves and regions
that are free from them. The boundary between such regions, 
the {\it chronology} horizon, is sometimes stable classically
but always quantum mechanically unstable. However, the
quantum instability of the Cauchy horizon still standing, 
the existence of solutions to the classical Einstein 
equations with stable Cauchy horizons remains a disturbing issue.
%
%

\section{Comments}
\label{cmc-s130}

In a classical setting, we can see that both linear and non-linear
examinations of the interior of black hole-de Sitter spacetimes
strongly suggest that the Cauchy horizon is stable for a finite,
but non-zero, measure on the space of black hole parameters 
$(M,Q,J,\Lambda)$. Moreover, simple but effective studies of the
stability issues~\cite{cmc-b160} yield intuitively compelling, and
pleasing, insights in to the mechanism at the heart of the Cauchy
horizon stability condition, Eq.~(\ref{cmc-e360}).
Whilst the majority of researchers within the field are confident of
the results and conclusions of the analyses in Sec.~\ref{cmc-s60},
some doubts, concerning the validity of the linear perturbation
studies, have been raised. It is worthwhile devoting some time
to a discussion of these doubts, addressing them directly with the
results and conclusions of Sec.~\ref{cmc-s60} at hand.

In Sec.~\ref{cmc-s70} we briefly discussed the concept of a linear
perturbation. For simplicity we shall consider a perturbation
described by a scalar field. We can expand the scalar field as
	\begin{equation}
	  \phi = \phi_{0} + \sum^{\infty}_{n=1}  \epsilon^{n}
	  \phi_{n} \; ,
	  \label{cmc-e385}
	\end{equation}
where $\phi_{0}$ is the background value of the field and 
$\phi^{n}$ represents the $n$-th order perturbation. 
Though not necessary here, we have introduced a dimensionless
parameter $\epsilon$  which, in calculations, keeps track
of the order of the perturbation. In the case of black holes the
background field is, in general, zero. For linear perturbation
studies one initially assumes that the scalar field is
{\it sufficiently weak }, in the sense that the stress-energy
associated with the field is negligibly small (no backreaction.)
Formally this implies we consider the field equations to
linear order in $\epsilon$. The stress-energy tensor
associated with the perturbation is 
quadratic in the field, and hence epsilon, so that to linear
order the stress-energy is ignored.
The spacetime is thus unaffected by the presence of the field and the
field evolution takes place on a fixed background spacetime.
If at any point of the evolution, the field becomes {\it large},
in the sense that its stress-energy becomes appreciable, then 
the linear theory is no longer valid and one must contemplate 
the construction of a model that takes account of non-linear
effects. Indeed this is exactly what happens at the Cauchy
horizon in the Reissner-Nordstr\"{o}m spacetime~\cite{cmc-b120}.
The stress-energy tensor of the field becomes increasingly large
as one approaches the Cauchy horizon. On the other hand, if the 
field remains {\it weak} everywhere in the region of interest then
the linear theory, and its results, remain valid. As we discussed
in the previous section, this is exactly what happens in the
Reissner-Nordstr\"{o}m-de Sitter spacetime. For $\kappa_{1}
> \kappa_{3}$ the fields stress-energy remains negligible and
the results, conventionally, are taken to be  representative of
the full theory. Of course, for $\kappa_{1} < \kappa_{3}$ the
theory breaks down and is qualitatively similar to the case 
$\Lambda = 0$. 

The doubts voiced about the results and conclusions of linear 
perturbation analyses concern 
themselves with precisely this conventional wisdom, that if a linear
theory is finite then the full theory is finite too.
Now, Eq.~(\ref{cmc-e385}) involves an infinite sum over the
field perturbations and although linear analyses show that
$\phi_{1}$ and its derivatives (appearing in the
stress-tensor) are well behaved at the Cauchy horizon
(for $\Lambda \neq 0$) this by no means guarantees
that the infinite sum converges to a finite value~\cite{cmc-b390}.
In fact, not only should this sum be convergent of course but also those
that appear in the stress-energy tensor.
To prove that these sums converge to a finite value would be
quite an undertaking, requiring a knowledge of the field
behavior to all orders. A direct approach to the problem,
showing convergence of the sums, is likely an impossible
task. One possible approach, which would not directly prove
convergence but could  at least  provide evidence of
convergence, is based on an approach initially used
by Ori to study non-linear perturbations in the Kerr
spacetime~\cite{cmc-b475}. This approach has been used to study the
non-linear effects of a scalar field in the 
Reissner-Nordstr\"{o}m-de Sitter spacetime~\cite{cmc-b400},
and predicts the correct linear behavior and allows one
to obtain the behavior of each term in the expansion described
by Eq.~(\ref{cmc-e385}). By comparing
the behavior of successive terms in the expansion, it may be possible 
to establish enough evidence to support the convergence
of the sum. However, the work of Brady {\it et al.}~\cite{cmc-b230}
has, to some extent, answered this question already. Indeed,
one can view this analysis as an alternative approach to
addressing the convergence problem by directly attempting
a non-linear study. That the results of this non-linear
analysis agree with the results from the linear studies
suggest that the conventional wisdom on linear perturbation
theory is valid. This idea can  even be taken one step further.
Instead of assuming a model where null dust mimics the
perturbations generated in the collapse, why not actually
study the gravitational collapse of a body, which eventually
forms a black hole? In this case one could attempt to follow
the evolution of the collapse through the event horizon  and in
to the interior, paying special attention to spacetime
near the Cauchy horizon. In the next section we shall address
this idea in more detail.
%
%

\section{Current Progress}
\label{cmc-s140}

One of the key ingredients to studying the interior of
any black hole spacetime  is a knowledge of the behavior
of fields crossing the event horizon, due to scattering in the
exterior of the black.  In a realistic collapse situation
these fields would always be present, as any initial perturbation
(or any perturbation that forms during the collapse) in the 
collapsing body becomes dynamical after the onset of collapse 
and thus emits gravitational or electromagnetic waves.
The waves, initially outgoing from the objects surface, will
scatter due to their interaction with the spacetime curvature
and a fraction of the wave will be reflected back to the
surface. If an event horizon forms, there will always be a 
scattered component of the wave that crosses the event horizon
and propagates through to the interior.

\subsection{Radiative Tails}

For black holes residing in asymptotically flat spacetimes,
the behavior of these radiating fields at sufficiently
late times after the collapse is well known now, both from
analytic calculations~\cite{cmc-b270,cmc-b280}  and from numerical
studies of collapse~\cite{cmc-b410,cmc-b420}.
The analytic and numerical studies both agree, that at late
times $(v \rightarrow \infty)$, along the event horizon, a 
physical field $\Psi$ decays according to
	\begin{equation}
	  \Psi \sim v^{-\gamma} \ \ \ , \ \ \ 
	  \gamma = 2\ell + 2 + P
	  \label{cmc-e490}
	\end{equation}
where $v$ is the standard advanced time coordinate of Sec.~\ref{cmc-s30},
$\ell$ is the multipole moment of the field $(\ell \geq s)$ with spin $s$ 
and P is a constant that assumes the value $0$ if there is an
initially static perturbation of the star and $1$ otherwise.
It is this late time decay that allows the gravitational field to relax 
to its final asymptotic state, parameterized by the three parameters 
associated with the non-radiatable multipoles $(\ell < s)$ of
the electromagnetic $(Q)$ and gravitational fields $(M,J)$ of the black
hole. These late time fields  are called the {\it tails} or {\it radiative
tails} of collapse. Their importance lies in their role as initial
data for studies of the interior, both analytically~\cite{cmc-b190}
and numerically~\cite{cmc-b430,cmc-b440}.  In the analytic studies
the late time behavior of the field is imposed as a restriction on the
stress-energy tensor of the matter that is flowing in to the hole across 
the event horizon (i.e the field is supposed to mimic the late time
component of the field scattered in to the black hole). 

For pure inflow models we have seen that the stress-energy has the form
given by Eq.~(\ref{cmc-e380}). The energy density measured by a freely
falling radial observer, crossing the Cauchy horizon in an
asymptotically flat black hole, is [Eq.~(\ref{cmc-e390})]
	\begin{equation}
	  \rho_{\scriptscriptstyle CH} = 
	  \frac{| E^{2} | }{4 \pi r^{2}_{\scriptscriptstyle CH}}
	  L(v) e^{2 \kappa_{\scriptscriptstyle CH} v} \; ,
	  \label{cmc-e500}
	\end{equation}
where to avoid confusion with the de Sitter case we have let 
$r=r_{\scriptscriptstyle CH}$ denote the location of the Cauchy 
horizon and $\kappa_{\scriptscriptstyle CH}$ its
surface gravity. Whereas in the de Sitter case the form of the luminosity
function $L(v)$ followed the requirement of a finite but non-vanishing flux
at the cosmological horizon, the situation in asymptotically flat
spacetimes is different. The requirement that the flux as measured by an
observer out at infinity approaches zero as $v$ tends to infinity, just
implies that $L(v)$ should tend to zero there. The form of $L(v)$ does not,
therefore, follow naturally. To ascertain the form of $L(v)$ one examines
the rate of change of the field, as measured by a freely falling observer
	\begin{equation}
	  \dot{\Psi} \equiv \frac{d \Psi}{d \tau} = \Psi_{, \alpha} u^{\alpha}
	  \; ,
	  \label{cmc-e510}
	\end{equation}
where $\tau$ is the observer's proper time. Now, close to the 
Cauchy horizon the four-velocity of the observer is dominated 
by the $u^{v}$ component, $\dot{v}$. Therefore, as $v$ tends to
infinity
	\[
	  \dot{\Psi} \sim \Psi_{,v} \dot{v} 
	  \sim v^{-\gamma-1} 
	  e^{\kappa_{\scriptscriptstyle CH} v} \; .
	\]
The flux $\rho_{\scriptscriptstyle CH}$ is proportional to 
$(\dot{\Psi})^{2}$, from which we can deduce that 
	\begin{equation}
	  L(v) \sim \alpha v^{-2(\gamma+1)}
	  \ \ \ \ {\rm as } \ \ \ \ v \rightarrow \infty \; ,
	  \label{cmc-e520}
	\end{equation}
where $\alpha$ is a constant.

To date, no analytic calculation of radiative tails in 
Reissner-Nordstr\"{o}m-de Sitter or Schwarzschild-de Sitter exists, but
it is surprising that the Brady-Poisson analysis did not actually require
us to know any details of the form of the tails near either the black hole
event horizon or the cosmological event horizon. Imposing a sufficiently
general requirement on the observed flux at the cosmological horizon 
(finite and non-zero) allows one to ascertain [See Sec.~\ref{cmc-s80},
Eqn.~(\ref{cmc-e410})] that
	\begin{equation}
	  L(v) \sim K(v) e^{-2 \kappa_{1} v} \; ,
	  \label{cmc-e530}
	\end{equation}
where $K(v)$ is some slowly varying function of $v$ that tends to a
finite, non-zero value, as $v$ tends to infinity. Reversing the argument
above for this form of the luminosity function we get that
	\begin{equation}
	  \Psi \sim e^{-\kappa_{1} v} 
	  \label{cmc-e540}
	\end{equation}
at late times, near the cosmological event horizon. 
Thus, one would claim that the
radiative tails in black hole-de Sitter spacetimes are exponential
with a folding time given by the surface gravity at the cosmological
event horizon. 

\subsection{Radiative Tails In Black Hole-de Sitter Spacetimes}

\label{cmc-s150}

We commented above that, unlike the case of black holes in asymptotically
flat spacetimes, no analytic work on radiative tails in black hole-de
Sitter spacetimes exists. This is either due to a lack of interest,
or more likely the comparative difficulty of working in such
spacetimes. Even in the case of Schwarzschild-de Sitter the analytic
work rapidly becomes difficult and tedious. The majority of the 
difficulties can be traced to the fact that
	\begin{equation}
 	  g^{\alpha \beta} \nabla_{\alpha} r \nabla_{\beta} r
	  = f(r) = 1 - \frac{2M}{r}+\frac{Q^{2}}{r^{2}}
	  -\frac{r^2}{\alpha^{2}} \; 
	  \label{cmc-e550}
	\end{equation}
has four roots and as $r$ tends to infinity 
	\[
	  f(r) \rightarrow -r^{2} / \alpha^{2} \; .
	\]
The lack of any detailed analysis, combined
with the difficulties of attempting an analytic investigation, led
Brady, Chambers, Krivan and Laguna~\cite{cmc-b450} to perform
the first numerical investigation in to the late time behavior of
fields propagating in black hole-de Sitter spacetimes.

\subsection*{Brady, Chambers, Krivan and Laguna 1997}

Brady {\it et al.} have studied, in some detail, the behavior of a
massless, minimally coupled scalar field  propagating on
spherically symmetric spacetimes with a positive cosmological
constant. Particular attention was focused on the late time
behavior of the fields in three particular regions;
(a)  the cosmological event horizon,
(b)  the black hole event horizon and 
(c)  future timelike infinity (point $D$ in Fig.~\ref{cmc-f30}) 
-- approached along surfaces of constant $r$
between these two horizons. The methods they employ
are similar to those used by Gundlach, Pullin and Price~\cite{cmc-b410}
in their numerical studies of radiative tails in asymptotically flat
black hole spacetimes,
	\begin{description}
	  \item[Linear Method]
	  The field propagates on the fixed background spacetimes of
	  \begin{itemize}
	    \item Schwarzschild-de Sitter
	    \item Reissner-Nordstr\"{o}m-de Sitter
	  \end{itemize}
	  \item[Non-Linear Method] 
	  The field is coupled to a general spherically
	  symmetric spacetime through the 
	  Einstein-Klein-Gordon field equations
	\end{description}

\subsection*{Linear Analysis}

The idea behind a linear analysis is similar to that discussed in 
the considerations of linear perturbation theory. One assumes that
the field is sufficiently weak that its effect upon the spacetime
is negligible. This is tantamount to assuming that the scalar field
is a linear perturbation on the spacetime, so that the stress-energy
tensor which, for the field under attention, is
	\begin{equation}
	  T_{\alpha \beta} = \phi_{,\alpha} \phi_{,\beta}
	  -\frac{1}{2} g_{\alpha \beta} \phi_{,\gamma}
	  \phi^{,\gamma}  \; ,
	  \label{cmc-e560}
	\end{equation}
is second order in the field and thus vanishes to linear order.
Linear analyses are not just favorable for their mathematical
simplicity. In cases where the spacetime curvature is small, as in 
the exterior region of a black hole, the results from linear
analyses are fairly representative of the results from non-linear
analyses as we shall demonstrate later. Indeed 
Price's original work on tails
in the  Schwarzschild spacetime, a linear perturbation
analysis, has been verified numerically both by linear and non-linear
evolutions~\cite{cmc-b410,cmc-b420} to a high degree of accuracy.

In terms of the standard advanced and retarded times $(u,v)$ 
the metrics for Schwarzschild-de Sitter and
Reissner-Nordstr\"{o}m-de Sitter are described by Eq.~(\ref{cmc-e150}),
	\[
	  ds^{2} = -f du dv + r^{2} d \Omega^{2} \; ,
        \]
with $f(r)$ given by Eq.~(\ref{cmc-e20}), from which
Schwarzschild-de Sitter is obtained simply by setting $Q=0$.  
On this background, the massless
minimally coupled scalar wave equation $\Box \phi = 0$ becomes
	\begin{equation}
	  \Psi_{,uv} = -\frac{1}{4} V_{\ell}(r) \Psi \; ,
	  \label{cmc-e580}
	\end{equation}
where the field has been decomposed in to spherical harmonics 
$\phi = \sum_{\ell, m} \Psi(u,v) Y_{\ell m}(\theta, \psi) r^{-1} $.
The function $V_{\ell}(r)$ is the effective
potential for the scalar field (Sec.~\ref{cmc-s70}) and has 
the following form
	\begin{equation}
	  V_{\ell}(r) = f(r) \left( \frac{\ell (\ell+1)}{r^{2}}
	  +\frac{f'(r)}{r^{2}} \right) \; ,
	  \label{cmc-590}
	\end{equation}
where $'$ denotes derivatives with respect to the function's argument.
Plots of the potential, between the cosmological and black hole 
event horizons, are given in Fig.~\ref{cmc-f70}.
%
%
\begin{figure}[t]
\leavevmode
\begin{center}
\epsfxsize=0.4\textwidth
\leavevmode\epsffile{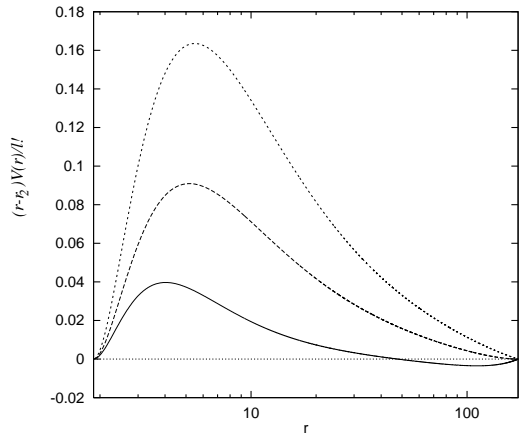}
\end{center}
\vspace{0.5cm}
\caption{The effective potential experienced by a scalar
field propagating on the fixed background spacetime of
a Reissner-Nordstr\"{o}m-de Sitter black hole with $Q=0.5$, 
$M=1$ and $\Lambda = 10^{-4}$. Shown are the potentials for the
$\ell = 0$ (solid), $\ell = 1$ (dashed) and the $\ell = 2$ 
(dotted) $\ell$-pole moments. The potential has been
scaled by $(r-r_{2}) / \ell !$ to accentuate the nature of the
potential when $\ell =0$. Unlike the higher $\ell$ modes, 
$V_{\ell = 0}$ shows a barrier followed by a well. For any
$\ell$-pole moment, the potential falls of exponentially 
as $r$ approaches either horizon.}
\label{cmc-f70}
\end{figure}
Equation~(\ref{cmc-e580}) is solved numerically by integrating it on the null 
grid defined by $u$ and $v$. Initial data (characteristic) is given 
by the value  of the
field $\Psi$ along two initial null surfaces $u=u_{0}$ and $v=v_{0}$.
The details of the numerical method are adequately described 
in~\cite{cmc-b410}. The results of the numerical integration are shown in 
Figs.~\ref{cmc-f80},~\ref{cmc-f90},~\ref{cmc-f100} and ~\ref{cmc-f110}.
%
%
\begin{figure}[t]
\leavevmode
\begin{center}
\epsfxsize=0.4\textwidth
\leavevmode\epsffile{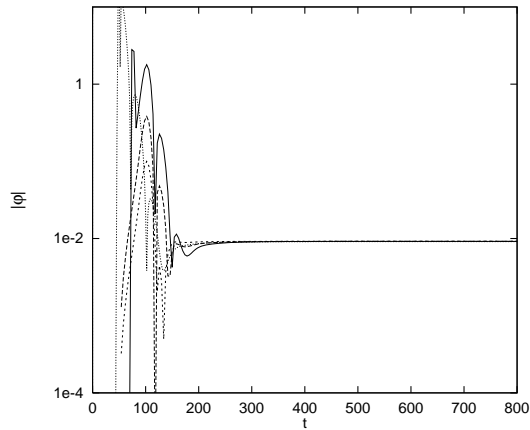}
\end{center}
\vspace{0.5cm}
\caption{A plot of $| \phi_{\ell = 0} |$ versus time for a 
Schwarzschild-de Sitter black hole spacetime. The field
behavior is shown along four different surfaces (see text).}
\label{cmc-f80}
\end{figure}
The initial data used to generate these figures was
	\begin{eqnarray*}
	  \Psi (u=0,v) &=& \exp \left[ - \frac{(v-v_{1})^2}{\sigma^{2}}
	  \right] \\
	  \Psi (u,v=0) &=& \Psi (u=0, v=0) \; ,
	\end{eqnarray*}
being representative of the other data sets employed by Brady 
{\it et al.}. All the graphs shown are for black holes with
their mass $(M)$ scaled to unity and $\Lambda = 10^{-4}$. Though
this choice of $\Lambda$ is arbitrary the results are 
qualitatively similar for any other value of $\Lambda$ that
is non-zero. The fields are shown plotted along four 
different surfaces,
	\begin{itemize}
	  \item
  	    The cosmological event horizon
	  \item
	    The black hole event horizon
	  \item
	    Two surfaces of constant $r$ approaching future
	    timelike infinity
	\end{itemize}
Since the late time behavior is identical along each surface
we shall not distinguish between them in the figures.
Figure~\ref{cmc-f80} displays the monopole field behavior
for a Schwarzschild-de Sitter black hole. At early times 
$( 0 < t < 200 )$ the field behavior is dominated by 
quasi-normal ringing, associated with complex characteristic
frequencies of the hole. At late times $ t > 200 $ the 
field approaches the same constant value on all four surfaces.
%
%
\begin{figure}[t]
\leavevmode
\begin{center}
\epsfxsize=0.4\textwidth
\leavevmode\epsffile{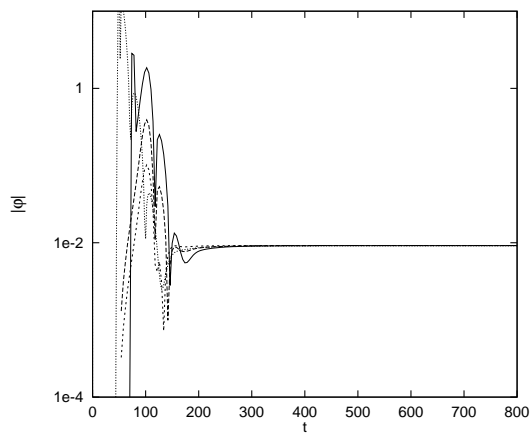}
\end{center}
\vspace{0.5cm}
\caption{ A plot of $| \phi_{\ell =0} |$ for a
Reissner-Nordstr\"{o}m-de Sitter black hole with
$Q=0.5$. The field behavior is shown along
four different surfaces (see text).}
\label{cmc-f90}
\end{figure}
Figure~\ref{cmc-f90} shows the same results for a 
Reissner-Nordstr\"{o}m-de Sitter black hole with $Q=0.5$. 
A period of quasi-normal ringing is followed by a relaxation
of the field to a constant value on all four surfaces. A detailed
investigation of the field's late time behavior reveals
	\begin{equation}
	  \phi_{\ell = 0} \simeq \phi_{0} 
	  + \phi_{1}(r) e^{-2 \kappa_{1} t} \; ,
	  \label{cmc-e600}
	\end{equation}
and that the constant field term $\phi_{0}$ scales like 
$\Lambda$~\cite{cmc-b450}. 
%
%
\begin{figure}[t]
\leavevmode
\begin{center}
\epsfxsize=0.4\textwidth
\leavevmode\epsffile{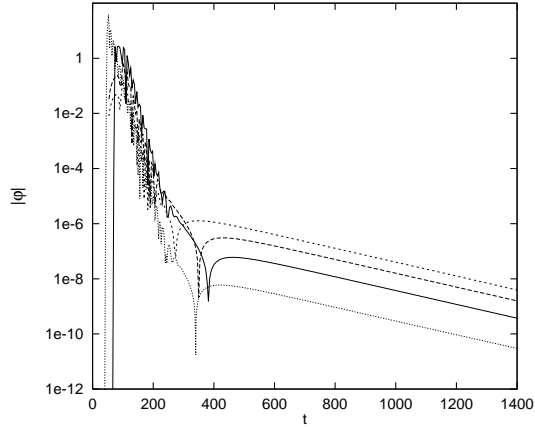}
\end{center}
\vspace{0.5cm}
\caption{A plot of $| \phi_{\ell = 1} |$ for a
Reissner-Nordstr\"{o}m-de Sitter black hole for
$Q=0.5$. The field is behavior is shown along
four different surfaces (see text).}
\label{cmc-f100}
\end{figure}
Figures~\ref{cmc-f100} and~\ref{cmc-f110} 
display the field behavior for $\ell = 1$ and the
$\ell = 2$ modes for the case $Q=0.5$. Again there is
a period of quasi-normal ringing followed by a
distinct exponential fall off. In general, an $\ell$-pole
mode of the field decays, at late times, like
	\begin{equation}
	  \phi_{\ell} \sim e^{- \ell \kappa_{1} t}
	  \ \  \ \ (\ell > 0) \; .
	  \label{cmc-e610}
	\end{equation}

\subsection*{Non-Linear Evolution}

If the scalar field is allowed to couple to the spacetime
via the stress-energy tensor, then  one must contend with 
the Einstein-Klein-Gordon field equations
	\begin{equation}
	 G_{\alpha \beta} \equiv R_{\alpha \beta} -
	 \frac{1}{2} g_{\alpha \beta} R = 8 \pi
	 T_{\alpha \beta} -g_{\alpha \beta} \Lambda \; ,
	 \label{cmc-e620}
	\end{equation}
where $T_{\alpha \beta}$ is the stress-energy tensor for
a massless, minimally coupled scalar field Eq.~(\ref{cmc-e560}).
For simplicity and tractability, attention is focused on
spherically symmetric spacetimes, whose line element can
be written as
	\begin{equation}
	  ds^{2} = -g \bar{g} du^{2} - 2 g du dr + r^2 d\Omega^{2}
	  \; .
	  \label{cmc-e630}
	\end{equation}
The Einstein field equations, Eq.~(\ref{cmc-e620}), then reduce to
	\begin{eqnarray}
	  ( \ln g)_{,r} &=&  4 \pi r^{-1} (h -\bar{h})^{2} \; ,
	  \label{cmc-e640} \\
	  (r \bar{g})_{,r} &=& g(1 -\Lambda r^{2})  \; ,
	  \label{cmc-e650} \\
	  (r \bar{h})_{,r} &=& h
	  \label{cmc-e660}
	\end{eqnarray}
and the scalar wave equation, $\Box \phi = 0$, becomes
	\begin{equation}
	  h_{,u}-\frac{\bar{g}}{2} h_{,r} =
	  \frac{(h-\bar{h})}{2 r}\left[
	  g(1-\Lambda r^{2}) - \bar{g} \right] 
	  \; ,
	  \label{cmc-e670}
	\end{equation}
where the auxiliary fields $(h,\bar{h} )$ are defined by
	\begin{equation}
  	  \bar{h} = \frac{1}{r} \int h dr \equiv \phi \; .
	  \label{cmc-e680}
	\end{equation}
Goldwirth and Piran~\cite{cmc-b460}
have devised a simple, but effective, numerical 
algorithm for  integrating these equations on a $(u,r)$ grid.
This method has been implemented by Gundlach {\it et al.} to 
study radiative tails in asymptotically flat black hole 
spacetimes~\cite{cmc-b410}. A refinement of this algorithm, which
reduces  numerical error near the $r=0$ origin has been
supplied by Garfinkle~\cite{cmc-b470} and used by Brady
{\it et al.} to integrate Eqs.~(\ref{cmc-e640})--(\ref{cmc-e670}).
Initial data for the problem is given by the value of the field along some
initial null cone centered at the origin of coordinates. Details 
of the numerical method can be found in~\cite{cmc-b410} and
references therein.
%
%
\begin{figure}[t]
\leavevmode
\begin{center}
\epsfxsize=0.4\textwidth
\leavevmode\epsffile{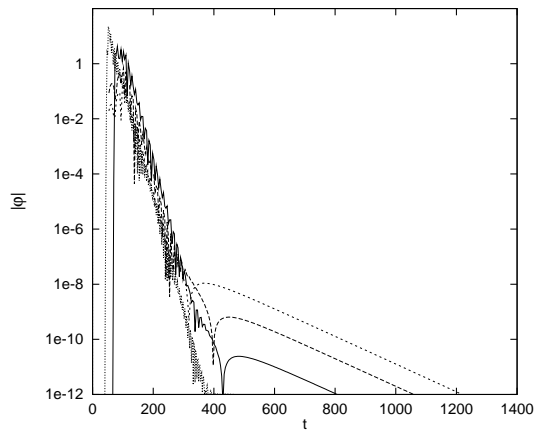}
\end{center}
\vspace{0.5cm}
\caption{A plot of $| \phi_{\ell = 2} |$ for a
Reissner-Nordstr\"{o}m-de Sitter black hole for $Q=0.5$.
The behavior of the field is shown along four surfaces (see text).}
\label{cmc-f110}
\end{figure}
In the graphs that follow, the initial data was Gaussian with
	\begin{equation}
	  \phi = \phi_{A} \left( \frac{r}{r_{0}} \right)^{2}
	  \exp \left[ -\frac{(r-r_{0})^{2}}{\sigma^{2}} 
	  \right] \; ,
	  \label{cmc-e690}
	\end{equation}
where $\phi_{A}$ is the amplitude of the field. The figures shown
are, again, for $\Lambda = 10^{-4}$. The behavior of the field 
in this case is along three surfaces,
	\begin{itemize}
	  \item
	    The cosmological event horizon
	  \item
	    The black hole event horizon
	  \item
	    A surface of constant $r$ approaching future timelike infinity
	\end{itemize}
Again, because the qualitative behavior of the field at late
times along each surface is similar, we do not explicitly
distinguish each surface in the figures. 
The restriction to spherical symmetry implies
that we gain only information about the $\ell = 0$ mode of
the field, so the results plotted are for $\ell =0$ only.
%
%
\begin{figure}[t]
\leavevmode
\begin{center}
\epsfxsize=0.4\textwidth
\leavevmode\epsffile{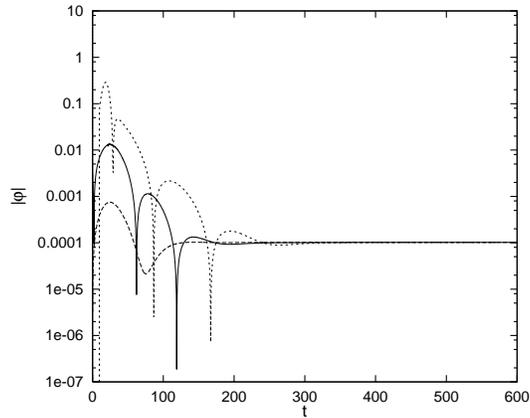}
\end{center}
\vspace{0.5cm}
\caption{A plot of $| \phi |$ versus time $t$  for
a non-linear evolution. The field behavior is plotted
for three the different surfaces explained in the
text}
\label{cmc-f120}
\end{figure}
In Fig.~\ref{cmc-f120} the field behavior can be seen to be
remarkably similar to that demonstrated by the linear analysis.
At late times the field approaches the same constant value along
all three surfaces, in confirmation of the test field results.
%
%
\begin{figure}[t]
\leavevmode
\begin{center}
\epsfxsize=0.4\textwidth
\leavevmode\epsffile{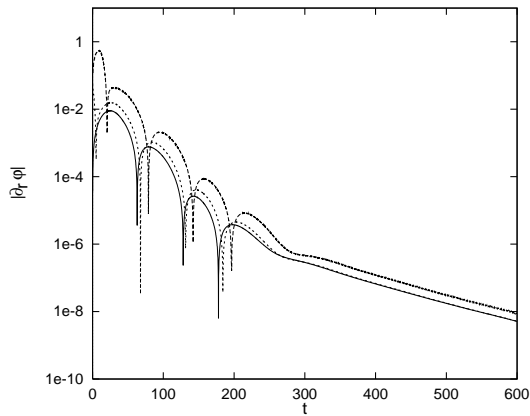}
\end{center}
\vspace{0.5cm}
\caption{A plot of the derivative of the field $| \phi_{,r} |$ 
versus time $t$ for
a non-linear evolution. The field behavior is plotted
for three the different surfaces explained in the
text}
\label{cmc-f130}
\end{figure}
Figure~\ref{cmc-f130} displays the behavior of $\phi_{,r}$, which
is proportional to $(\bar{h}-h)$ along surfaces of constant $r$.
Brady {\it et al.} find that at late times
	\begin{equation}
	  (\bar{h}-h) \sim e^{-2 \kappa_{1} t} 
	  \ \ \ \ {\rm as} \ \ \ \ u \rightarrow \infty \; ,
	  \label{cmc-e700}
	\end{equation}
so that 
	\begin{equation}
	  \phi_{\ell =0} \simeq
	  \phi_{0} + \phi_{1}(r) e^{-2 \kappa t}
	  \ \ \ \ {\rm as } \ \ \ \
	  t \rightarrow \infty \; ,
	\label{cmc-e710}
	\end{equation}
in agreement with the linear perturbation analysis.

\subsection{Conclusions}

For a massless, minimally coupled scalar field propagating
on the fixed backgrounds of Schwarzschild-de Sitter and
Reissner-Nordstr\"{o}m-de Sitter, the late time behavior 
along  the cosmological event horizon, the black hole event
horizon and  at future timelike infinity is given by
	\[
	  \phi \sim e^{-\ell \kappa_{1} t} \; ,
	\]
for $\ell > 0$.
For $\ell = 0$ behavior is slightly different, 
	\[
	  \phi_{\ell =0} \sim \phi_{0} + \phi_{1} 
	  e^{-2 \kappa_{1} t} \; ,
	\]
asymptoting to a constant value at late times, which is
ultimately confirmed by the non-linear evolution of a
spherically symmetric scalar field coupled to general 
relativity.

The behavior of the $\ell = 0$ mode is, in some ways,
unusual. As $t$ tends to infinity, the field
mode approaches a constant value, in fact the same
constant value, at both the cosmological
event horizon and the black hole event horizon. There
is no analogue of this for the case of black holes in flat
space. While Price's original work~\cite{cmc-b270} demonstrates
that there can be no static solutions to the scalar
wave equation that are well behaved at infinity and the
black hole event horizon, one can actually have the 
trivial static solution $\phi = constant$. The same holds
true in black hole-de Sitter spacetimes. Even though the 
constant solution carries no stress-energy, it is intriguing
that the appearance of a constant field value
at the horizons does not occur in the flat space examples.
An examination of the zero frequency reflection and transmission
coefficients in the region exterior to a de Sitter black hole, 
shows they are non-zero, allowing the propagation of a
constant mode to both horizons.  In this respect,
scattering in the exterior of a black hole-de Sitter
spacetime is somewhat similar to scattering in the interior
of a Reissner-Nordstr\"{o}m black hole. However, in the exterior 
of a Reissner-Nordstr\"{o}m black hole no such constant mode propagation is
allowed. It is likely that the constant mode propagation in an 
$\ell = 0$ mode and the anomalous dip in the effective potential for that 
mode (Fig.~\ref{cmc-f70}) are not a mere coincidence, though
no detailed investigation of this relation has been  forthcoming.
What is known is that for a conformally invariant scalar 
field the well in the potential disappears, and in this case
the constant field is zero.
A similar well in the effective potential occurs for the 
case of the gravitational perturbations detailed by Mellor and 
Moss~\cite{cmc-b100}. It seems reasonable to suspect similar
constant mode behavior will be observed in this case too~\cite{cmc-b471}.

At the start of this section on current work, we commented on how
the minimal requirement of Brady-Poisson implicitly imposes
form on the late time behavior of the perturbing field near
the cosmological event horizon, described by Eq.~(\ref{cmc-e540}).
Since the minimal requirement is essential to the stability
condition expressed in Eq.~(\ref{cmc-e360}), it is important
that the numerics support it.
The numerical results have shown that we can, in general, expand the
scalar field as
	\[
	  \phi(t,r) = \phi_{0}+\sum_{n=1}^{n=\infty} \phi_{n}(r)
	  e^{-n \kappa_{1} t} \; ,
	\]
in the exterior at late times.
Eq.~(\ref{cmc-e540}) gives the form of the field we should
expect from imposing the minimal requirement,
	\[
	  \phi \simeq a(r)   + b(r) e^{-\kappa_{1} t} \; .
	\]
We should note that this is a more general solution than that
proposed in Eq.~(\ref{cmc-e540}), where we dropped the constants
of integration for simplicity. Making the identifications
$ a(r) = \phi_{0}$ and $ b(r) = \phi_{1}(r)$ demonstrates that the
numerical results confirm the minimal requirement of Brady 
and Poisson. It is worth noting that it is the $n=1$, or
$\ell =1$ mode, that actually provides the required form for
the radiative tails. The contribution from the other $\ell$
modes leads to a vanishing flux at the Cauchy horizon. In
any generic perturbation we would expect all $\ell$-pole
moments to be present, thus  assuring the correctness of
the Brady-Poisson model.
%
%

\section{Discussion}
\label{cmc-s160}

We began this review by stressing the need for a
precise formulation and proof of the strong cosmic
censorship hypothesis. It is almost twenty years
since Penrose conjectured this stronger form of
cosmic censorship~\cite{cmc-b472}, and today it still remains
a cardinal, unsolved problem of general relativity.
Instead of seeking a correct formulation of
strong cosmic censorship, we have opted for a much 
simpler route -- a search for reasonable counter-examples,
in the belief that an understanding of these models will
inevitably lead to a deeper comprehension of
the censor issue. This hunt for counter-examples has lead 
us in to realm of black hole-de Sitter spacetimes and
to pose the question
	\begin{quote}
	  ``Are black holes immersed in de Sitter space
	    counter-examples to the current formulation
	    of strong cosmic censorship?"
        \end{quote}
The answer to this appears, at least classically, to
be yes. The linear perturbation analyses in 
Reissner-Nordstr\"{o}m-de Sitter and Kerr-Newman-de Sitter and 
the backreaction calculations, performed in
spherical symmetry, agree -- the Cauchy horizon 
in black hole-de Sitter spacetimes is stable provided that
	\[
	   \kappa_{1} > \kappa_{3} \; .
	\]
Quantum mechanically the answer appears to be no. It
wasn't obvious (to me at least) that this divergence couldn't
just be due to a bad choice of vacuum state. To prove
quantum instability one actually needs to prove there are
no quantum states that are regular on all three horizons.
Eric Poisson has kindly pointed out to me that although
Markovi\'{c} and Poisson did not give this proof, one
does exist, at least in the two dimensional case. The
proof is provided in Eric's contribution elsewhere in
these proceedings.
It might be interesting to see if this results is
easily continued to the four dimensional case.
Occasionally, the results of a two dimensional
calculation do not concur with the results
of a four dimensional analysis.
An excellent demonstration of this occurs in
the extreme Reissner-Nordstr\"{o}m spacetime. In two
dimensions it has been shown~\cite{cmc-b473} that stress-energy
of a quantized field will diverge on the event horizon of
an extreme black hole, whereas a four dimensional calculation
shows no sign of a divergence~\cite{cmc-b474}. It maybe that
a four dimensional calculation of quantum effects near
the Cauchy horizon will lead to a somewhat different conclusion
than that reached by the two dimensioanl analysis. The quantum
instability is undoubtedly an important question. However, the
fiery marriage between general relativity and quantum mechanics,
that is {\it Quantum-Gravity}, has yet to be consummated.
For that reason we have concentrated mainly on the classical
stability of the Cauchy horizon in black hole-de Sitter
spacetimes.  The existence of solutions to the classical field
equations  exhibiting stable Cauchy horizons faces 
classical physics with many problems, including  a description of the
singularity and how to attempt physics in the presence of
causality violating curves.

\subsection*{A look to the Future}
We now turn our attention to the future. The following comments
are based purely on personal speculation  about future work
and the role it will play in answering the question posed above.
The purpose of this section is not just to seek the truth, but to
try to encourage a more active participation in the study of
black hole-de Sitter spacetimes. 

With the results of Brady {\it et al.}, on the form of
radiative tails in Reissner-Nordstr\"{o}m-de Sitter~\cite{cmc-b450},
things are pretty much set up to allow a numerical
investigation of the interior. An adequate amount
of numerical algorithms and techniques now exist, which have been 
used in  the study of the Reissner-Nordstr\"{o}m 
spacetime~\cite{cmc-b420,cmc-b430,cmc-b440}, and
should allow this problem to be confronted with some ease.
It is hoped that the numerical results will provide further
evidence toward the conjecture that the Cauchy horizon
in black hole-de Sitter spacetimes can be stable for
a fixed, non-zero, region of its parameter space.
A numerical study will, at the least, allow us to
gain some insight in to the behavior of the spacetime
in the vicinity of the Cauchy horizon and the behavior of
fields propagating in the interior. These results
can be used to gauge the adequacy of the linear
analysis in describing the physics at the horizon and
could provide information to initiate new analytic approaches 
for studying the interior.

A preliminary investigation of the numerical problem suggests that
one should take an approach similar to that used by
Brady and Smith~\cite{cmc-b440} for studying the interior
but implement the algorithm of Burko and Ori~\cite{cmc-b440}.
In this case the initial data is specified along some
null surface that is taken to coincide with the event
horizon. The alternative is to specify generic (Gaussian) initial
data outside the hole (see~\cite{cmc-b430,cmc-b440}) and 
let it evolve, through the event, to the interior.
For Reissner-Nordstr\"{o}m this is quite adequate and
has the advantage that one does not have to know the
precise form of the radiative tails. For the de Sitter
case this is not really a satisfactory approach. As
we saw in Sec.~\ref{cmc-s150}, the restriction to
spherical symmetry [Eq.~(\ref{cmc-e630})] implies that
if we start from generic initial data outside the
hole then at the event horizon we will have only the
radiative tail of the $\ell = 0$ mode. The form of
this tail, Eq.~(\ref{cmc-e710}), does not satisfy
the minimal requirement of Brady and Poisson discussed
in Sec.~\ref{cmc-s80}, for this mode there is a
vanishing flux of energy at the cosmological
event horizon. To obtain a non-vanishing flux
requires the introduction of higher $\ell$-modes,
specifically the $\ell = 1$ mode. Specifying
the initial data along the event horizon, as
Brady and Smith did, allows,
in some respects, the added flexibility of incorporating
the effects of additional $\ell$-modes.

One of the biggest problems facing anyone attempting
to study the interior of a Reissner-Nordstr\"{o}m-de Sitter
black hole is the search for an initial data set that
evolves in to a solution with a stable Cauchy horizon.
From Fig.~\ref{cmc-f60}, it is easy to see that the
stability region is extremely small, in fact 0.6\%
of the entire parameter space by area. It might
be possible, instead of hunting for initial data,
to attempt some type of shooting method, with boundary
conditions at the future event horizon and  
cosmological horizon. At present this  is still
a suggestion and may prove to be a dead end.
Another approach which initially appeared sound was
to fix regularity conditions, suggested by
the analytic studies, on the spacetime and the fields 
at the Cauchy horizon and evolve them
backward in time to the event horizon. The philosophy
behind this idea being that one starts off with 
a solution that has a regular Cauchy horizon in order to see
whether or not it can evolve from regular data
at the event horizon. Of course we expect the data
corresponding to the field to
have the exponential form required by the tails. The downfall
of this approach is that derivatives of the field,
at the Cauchy horizon, vanish. It seems likely therefore
that the evolution will end up showing no field propagating
through the interior as a whole, i.e, vanishing field
perturbation at the event horizon. Currently this
remains the most difficult step on the path toward
a numerical investigation of the interior. While
the feeling is that this is not an insurmountable problem, 
it seems it will require some thought.

%
%

\appendix

\section{Differential Proper Times}
\label{cmc-a10}

We present the arguments of Brady and Poisson~\cite{cmc-b160}
which demonstrate that a necessary condition for Cauchy horizon
instability is that the  ratio of {\it differential} proper times
be divergent.

We consider two observers
in the spacetime, an external observer in region II approaching the 
cosmological event horizon and an internal observer in region III
approaching the Cauchy horizon. Each observer measures the proper
time, $\tau_{\scriptscriptstyle II}$ and $\tau_{\scriptscriptstyle III}$ 
respectively, between two
successive wavefronts, labelled $v_{0}$ and $v_{0}+dv$.
The situation is shown schematically in Fig.~\ref{cmc-fa10}.
%
%
\begin{figure}[t]
\leavevmode
\begin{center}
\epsfxsize=0.4\textwidth
\leavevmode\epsffile{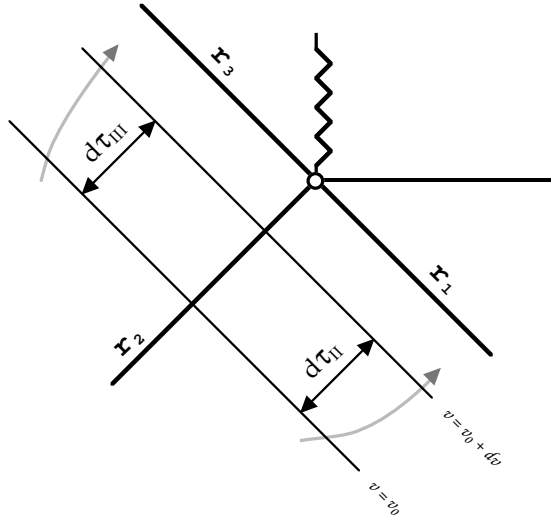}
\end{center}
\vspace{0.5cm}
\caption{A schematic representation of the blueshift vs.~redshift
argument of Poisson and Brady. Shown are
the inner horizon ($r_{3}$), the outer horizon ($r_{1}$)
and the black hole event horizon ($r_{2}$). Also
shown are the paths of the two observers (light arrows)
approaching their respective horizons. {\protect $v_{0}$}
and {\protect $v_{0} +dv$} represent two radially ingoing 
null rays, meant to represent two successive wave crests. 
The external observer measures a proper time 
{\protect $d\tau_{\scriptscriptstyle II}$} between the crests
whilst the internal observer measures a proper time
{\protect $d\tau_{\scriptscriptstyle III}$}}
\label{cmc-fa10}
\end{figure}
In region II, as the observer nears the cosmological
event horizon, the $u^{v}$ component of their four-velocity,
$\dot{v}$, approaches
	\begin{equation}
	  \dot{v}_{\scriptscriptstyle II} \equiv 
	  \frac{d v_{\scriptscriptstyle II}}{d 
	  \tau_{\scriptscriptstyle II}}
	  \simeq \frac{|E_{\scriptscriptstyle II}|}{r_{1} \kappa_{1}}
	  e^{\kappa_{1} v} \; ,
	\label{cmc-ea10}
	\end{equation}
where we have used the results of Sec.~\ref{cmc-s40} [Eq.~(\ref{cmc-e283})].
Similarly, for observers in region III approaching the Cauchy
horizon,
	\begin{equation}
	  \dot{v}_{\scriptscriptstyle III} \equiv 
	  \frac{d v_{\scriptscriptstyle III}}{d 
	  \tau_{\scriptscriptstyle III}}
	  \simeq \frac{|E_{\scriptscriptstyle III}|}{r_{3} \kappa_{3}}
	  e^{\kappa_{3} v} \; ,
	\label{cmc-ea20}
	\end{equation}
[See Eq.~(\ref{cmc-e286})]. Since the observers are measuring
proper time between the same successive wavefronts
($dv_{\scriptscriptstyle II} = dv_{\scriptscriptstyle III} \equiv dv$)
and because both horizons are located at the same advanced time, then
	\begin{equation}
	  \frac{\dot{v}_{\scriptscriptstyle II}}{
	  \dot{v}_{\scriptscriptstyle III}}
	  = \frac{d\tau_{\scriptscriptstyle III}}{d
	  \tau_{\scriptscriptstyle II}}
	  \sim e^{(\kappa_{1} - \kappa_{3}) v}
	  \ \ \ {\rm as} \ \ \ v \rightarrow \infty \; .
	\label{cmc-ea30}
	\end{equation}
Therefore, if $\kappa_{3} > \kappa_{1}$, the interior observer
measures an increasingly smaller proper time between successive wave crests
as she/he approaches the Cauchy horizon indicating a blueshift and
hence an instability of the Cauchy horizon. If, on the other hand
$\kappa_{1} > \kappa_{3}$, the observer measures an increasingly
larger proper time between the wavefronts as he/she approaches the
Cauchy horizon, indicating a redshift. In this case the Cauchy
horizon is stable. Even without performing the calculation, it
is not hard to see that these observers take a finite proper time
to reach there respective horizons. 
One can also see the inevitability of the blueshift
instability of the Cauchy horizon in the Reissner-Nordstr\"{o}m 
solution. Here, the cosmological event horizon is replaced by
by future null infinity,
${\cal J}^{+}$, and $\kappa_{1} \rightarrow 0$. In this case the internal
observer will always measure an vanishingly small proper time between
the wave crests as he approaches the inner horizon and so the 
Cauchy horizon is always unstable. This instability is
due to the infinite time compression effects we discussed
in the Introduction. Thus, we can conclude that
while the infinite compression of the ratio of
proper times is a sufficient
condition for Cauchy horizon stability, it is not a necessary condition.
A necessary condition is an infinite compression of the ratio
of {\it differential} proper times expressed in Eq.~(\ref{cmc-ea30}).

\section{Ori-Model} 
\label{cmc-a20}

The Poisson-Israel model of mass inflation~\cite{cmc-b190} models the 
fluxes of radiation in the interior, generated during 
the collapse to form a black hole,
as a crossflow of lightlike particles moving radially
outward and inward respectively. Whilst the form of the inflow is 
crucial to the analysis (modelling the late time behavior of the
fields crossing the event horizon) the nature of the outflux is largely
irrelevant, its presence is required only to precipitate a contraction
of the generators of the Cauchy horizon. In the Ori model the outflux
is modelled as a thin lightlike shell $\Sigma$, which allows an
exact mass inflation solution. The generalization of this model to
black holes in de Sitter space is simple~\cite{cmc-b230}.
%
%
\begin{figure}[t]
\leavevmode
\begin{center}
\epsfxsize=0.4\textwidth
\leavevmode\epsffile{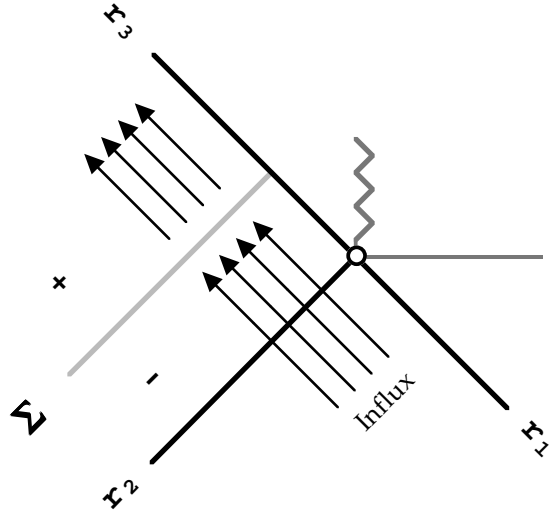}
\end{center}
\vspace{0.5cm}
\caption{The conformal diagram for the Ori model in
the Reissner-Nordstr\"{o}m-de Sitter spacetime. Shown 
are the locations of the cosmological event horizon $r_{1}$,
the black hole event horizon $r_{2}$ and the Cauchy
horizon $r_{3}$. Also shown is an influx of radiation
crossing the event horizon (arrows) and the outflow
of radiation, idealized as a thin lightlike shell $\Sigma$,
emanating from the collapsing star's surface to the left of
the diagram (not shown). The outflow naturally divides the 
spacetime up in to two distinct regions, that to its future $(+)$
and that to its past $(-)$.}
\label{cmc-fa20}
\end{figure}
Figure~\ref{cmc-fa20} depicts the conformal diagram of the situation.
The null shell $\Sigma$ divides the interior spacetime in to two
distinct regions, that to the past of $\Sigma$ which we label $(-)$ 
and that to the future, which we label $(+)$. On either side of the 
shell the spacetime is that of Reissner-Nordstr\"{o}m-Vaidya-de Sitter
(RNVDS), with metrics
	\begin{eqnarray}
	  ds^{2}_{\pm} &=& -f_{\pm} dv^{2}_{\pm} + 2 dv_{\pm} dr_{\pm}
	  +r^{2}_{\pm} d\Omega^{2}_{\pm} 
	  \label{cmc-eb10}              \\
	  f_{\pm} &=& 1 -\frac{2 m_{\pm}(v_{\pm})}{r_{\pm}}+
	  \frac{Q^{2}_{\pm}}{r^{2}_{\pm}}-\frac{r^{2}_{\pm}}{\alpha^{2}}
	  \; .
	  \label{cmc-eb20}
	\end{eqnarray}
The stress-energy tensor is
	\begin{equation}
	  T_{\alpha \beta}^{\pm} = \frac{L(v_{\pm})}{4 \pi r_{\pm}^{2}}
	  (\partial_{\alpha} v_{\pm}) (\partial_{\beta} v_{\pm}) \; .
	  \label{cmc-eb30}
	\end{equation}
The luminosity function $L(v)$ is related to the mass function $m(v)$ 
through the $G_{vv}$ component of the Einstein equations
	\begin{equation}
	  L(v) = \frac{d m(v)}{dv}
	  \label{cmc-eb35}
	\end{equation}
\subsection*{Matching}
The problem is to ascertain the nature of the spacetime to the future 
of $\Sigma$, the $(+)$ region. This is simply done by matching the
two RNVDS spacetimes across the shell. The first requirement is that 
metric tensor $g_{\alpha \beta}$ be continuous across $\Sigma$, which
requires that $r_{+} = r_{-} \equiv r$ and $d\Omega^{2}_{+}
= d\Omega^{2}_{-} \equiv d \Omega^{2}$. We also impose the physical
conditions that $\Sigma$ be electrically neutral, so that $Q_{+} =
Q_{-} \equiv Q$ and that it be pressureless, which requires that
$\lambda_{+} = \lambda_{-} \equiv \lambda$, where $\lambda$ is the affine
parameter along $\Sigma$. The two important matching conditions that
can be derived from these are
	\begin{itemize}
	  \item 
	  The null generators of $\Sigma$ are the same on either side
	    \begin{equation}
	      2 dr = f_{+} dv_{+} = f_{-} dv_{-} 
	      \label{cmc-eb40}
	    \end{equation}
	  \item
	  Flux continuity across $\Sigma$
	    \begin{equation}
	      \frac{1}{f_{+}} \frac{dm_{+}}{d \lambda} =
	      \frac{1}{f_{-}} \frac{d m_{-}}{d \lambda}
	      \label{cmc-eb50}
	    \end{equation}
	\end{itemize}
The first condition is given by the continuity of the metric tensor
and that null geodesics satisfy $dr/dv = f/2$. The second condition is
the requirement that $\lambda_{+} = \lambda_{-}$, which can be 
shown~\cite{cmc-b480} to be equivalent to the requirement that 
$T_{\alpha \beta}^{+} m^{\alpha}_{+} m^{\beta}_{+} = 
T^{-}_{\alpha \beta} m^{\alpha}_{-} m^{\beta}_{-}$, where 
the $m^{\alpha}_{\pm}$ are the tangents
to the null generators of $\Sigma$ on either side. Thus the 
requirement of pressureless implies that $\Sigma$ does not interact 
with the influx.
\subsection*{An Exact Solution}
In the $(-)$ region we have that
	\begin{equation}
  	  L(v_{-}) = \frac{dm_{-}}{dv_{-}} = K(v_{-}) 
	  e^{-2 \kappa_{1} v_{-}} \; .
	  \label{cmc-eb60}
	\end{equation}
Integrating this gives
	\begin{equation}
	  m_{-}(v_{-}) = M - \frac{\alpha}{2 \kappa_{1}}
	  e^{-2 \kappa_{1} v_{-}} \; ,
	  \label{cmc-eb70}
	\end{equation}
where for convenience we have assumed that $K(v_{-})$ is a
constant $\alpha$. In reality $K(v_{-})$ is a  slowly varying
function of $v_{-}$ and numerical analyses suggest that to
first order $K(v_{-})$ is constant. $M = m_{-}(\infty)$ is
the final asymptotic mass. Equation~(\ref{cmc-eb40})  gives that
the null geodesics in region $(-)$ obey
	\begin{equation}
	  \frac{dr}{dv_{-}} = \frac{f_{-}}{2} \; .
	  \label{cmc-eb80}
	\end{equation}
Expanding $f_{-}$ about $r=r_{3}$ and letting $v_{-}$ tend to
infinity allows us to integrate this expression 
	\begin{equation}
	  \kappa_{3} (r-r_{3}) \simeq \beta e^{-\kappa_{3} v_{-}}
	  +\frac{\alpha \kappa_{3}}{2 \kappa_{1} r_{3}} 
	  \frac{e^{-2 \kappa_{1} v_{-}}}{(\kappa_{3} - 2 \kappa_{1})}
	  \; ,
	  \label{cmc-eb90}
	\end{equation}
where $\beta$ is a constant of integration. Then we can write the 
asymptotic form of $f_{-}$
	\begin{equation}
	  f_{-}(r,v_{-}) \simeq -2 \beta e^{-2 \kappa_{3} v_{-}}-
	  \frac{2 \alpha}{r_{3} (\kappa_{3} -2 \kappa_{1})}
	  e^{-2 \kappa_{1} v_{-}} \; ,
	  \label{cmc-eb100}
	\end{equation}
as $r \rightarrow r_{3}$ and $v_{-} \rightarrow \infty$. The
second matching condition, Eq.~(\ref{cmc-eb60}),
allows us to obtain $m_{+}$ as a function
of $v_{-}$,
	\begin{equation}
	  \int \frac{d m_{+}}{f_{+}} = \int \frac{d m_{-}}{f_{-}}
	  =\int \frac{d m_{-}}{dv_{-}} \frac{1}{f_{-}} dv_{-} 
	  \; .
	  \label{cmc-eb110}
	\end{equation}
Using Eqs.~(\ref{cmc-eb20},~\ref{cmc-eb70}) and~(\ref{cmc-eb100})
it is easy to show that 
	\begin{equation}
	  m_{+}(v_{-}) \simeq M + \frac{\gamma \beta M (\kappa_{3}
	  -2 \kappa_{1})}{\alpha} + \gamma M 
	  e^{(\kappa_{3} -2 \kappa_{1}) v_{-}}
	  \; ,
	  \label{cmc-eb120}
	\end{equation}
where $\gamma$ is another constant of integration. We can see immediately
from this relation that the mass function to the future of the shell
will inflate if and only if $\kappa_{3} > 2 \kappa_{1}$ which agrees
with the work by Brady {\it et al.}~\cite{cmc-b230}. We shall concentrate
on the case that $\kappa_{3} > 2 \kappa_{1}$, the other cases follow
the same line of reasoning. We can therefore estimate the leading
behavior of the mass to be
	\begin{equation}
	  m_{+}(v_{-}) \sim \gamma M e^{(\kappa_{3} - 2 \kappa_{1}) v_{-}}
	\ \ \ \ {\rm as } \ \ \ \ v_{-} \rightarrow  \infty \; .
	  \label{cmc-eb130}
	\end{equation}
We require to know how $v_{+}$ and $v_{-}$ are related, which we
can do by using Eq.~(\ref{cmc-eb40})
	\begin{equation}
 	  \int dv_{+} = \int \frac{f_{-}}{f_{+}} dv_{-} \; .
	  \label{cmc-eb140}
	\end{equation}
Using Eqs.~(\ref{cmc-eb20},~\ref{cmc-eb100}) and~(\ref{cmc-eb130})
we find that
	\begin{equation}
	  v_{+} \sim - \frac{\alpha}{\gamma M (\kappa_{3}-2\kappa_{1})
	  \kappa_{3}} e^{-\kappa_{3} v_{-}} \; ,
	  \label{cmc-eb150}
	\end{equation}
where, without loss of generality, we have set the constant of 
integration to zero. If we remember that the regular
coordinate at the Cauchy horizon is $V = -e^{-\kappa_{3} v_{-}}$
[Eq.~(\ref{cmc-e230})] we can see that $v_{+}$ is Kruskal-like,
	\begin{equation}
	   v_{+} \sim \frac{\alpha}{ \gamma M (\kappa_{3} -
	   2 \kappa_{1}) \kappa_{3}} V \; ,
	   \label{cmc-eb160}
	\end{equation}
then
	\begin{equation}
	  v_{-} \sim  - \frac{1}{\kappa_{3}}\left| -V \right|
	  \; .
	  \label{cmc-eb170}
	\end{equation}
The mass function $m_{+}(V)$ thus inflates as
	\begin{equation}
	  m_{+}(V) \sim \gamma M (-V)^{\frac{2 \kappa_{1} - \kappa_{3}}{
	  \kappa_{3}}} \ \ {\rm as } \ \
	  V \rightarrow 0 \; .
	  \label{cmc-eb180}
	\end{equation}
We define a new coordinate $U$ such that
	\begin{equation}
	  - \delta dU = m_{+} dv_{+} + r dr 
	  \ \ {\rm where } \ \
	  \delta =\frac{\gamma M (\kappa_{3} -2 \kappa_{1}) r_{3}}{
	  \alpha \kappa_{3}} \; '
	  \label{cmc-eb190}
	\end{equation}
so that close to the Cauchy horizon the metric becomes
	\begin{equation}
  	  ds^{2}_{+} = -2 e^{2 \sigma} dU dV + r^{2} d\Omega^{2}
	  \ \ {\rm where } \ \
	  e^{2 \sigma} =\frac{1}{\kappa_{3}^{2}} \frac{r_{3}}{r}
	  \label{cmc-eb200}
	\end{equation}
From Eq.~(\ref{cmc-eb190}) we can obtain the behavior of the radius
	\begin{equation}
	  r^{2} = r_{3}^{2}-2 U \delta + \frac{ \alpha}{(\kappa_{3}
	  -2 \kappa_{1}) \kappa_{1}} (-V)^{2 \frac{\kappa_{1}}{
	  \kappa_{3}}} \; ,
	  \label{cmc-eb210}
	\end{equation}
which reflects the slow contraction of the Cauchy horizon.

%
%


%
%
%

\begin{thebibliography}{99}
\bibitem{cmc-b01}
G.T. Horowitz and H.J. Sheinblatt, Phys. Rev. D {\bf 55}, 650 (1997).
\bibitem{cmc-b02}
Our notaion for the causal structure follows Ref.~\cite{cmc-b20},
Chap.~8, p.~188.
\bibitem{cmc-b03}
The phrase ``terminating it at the Cauchy horizon" is
probably a little strong. There are indications that the
horizon is traversible. However, this point was the focus
of much debate during the Haifa meeting and it is clear that
not everybody agrees with the interpretations presented in 
favor of traversability.  The issue is an important one and is
adequately dealt with elsewhere in these preceedings, where
the reader can make up his or her own mind.
\bibitem{cmc-b04}
A brief search of the astro-ph archive at lanl, for the year 1997 to date,
revealed 6 submissions offering evidence that challenges the 
`establishment' view of a vanishing cosmological constant.
\bibitem{cmc-b05}
R. Herman and W.A. Hiscock, Phys. Rev. D {\bf 46}, 1863 (1992).
\bibitem{cmc-b06}
This is only true in a classical context. We shall see later  that
quantum mechanically, the Cauchy horizon in black hole-de Sitter
spacetimes appears to be unstable.
\bibitem{cmc-b10}
B. Carter, in {\it Black Holes}, edited by C.M. De Witt and B.S.
De Witt (Gordon and Breach, New York, 1973).
\bibitem{cmc-b20}
R.M. Wald, {\it General Relativity}, (The University of Chicago Press,
Chicago and London, 1984).
\bibitem{cmc-b30}
By {\it similar} we mean a spacetime for which $g^{\alpha \beta}
\nabla_{\alpha} r \nabla_{\beta} r \equiv f(r)$.
\bibitem{cmc-b40}
In this and what follows in subsequent sections, we shall implicitly
assume that there exist  four distinct roots to Eq.~(\ref{cmc-e70}),
and not concern ourselves with solutions possessing coincident horizons. 
\bibitem{cmc-b50}
For the Kerr-Newman-de Sitter spacetime the root $r_{4}$, though negative,
does correspond to a physical horizon lying beyond the  
ring singularity at $r=0$.
\bibitem{cmc-b60}
R. Penrose, in {\it Relativity, Groups and Topology}, edited
by C.M. De Witt and B.S. De Witt (Gordon and Breach, New York, 1963).
\bibitem{cmc-b70}
This point assumes the collapse becomes dynamical in region II.
\bibitem{cmc-b80}
The definitions of $u$ and $v$ in region II are not unique either,
but they do follow the more usual convention. The main reason for
introducing a less conventional definition of $u$ and $v$ here is to
alert the reader to the fact it is possible and, in certain circumstances,
it can be advantageous to define $u$ and $v$  differently.
\bibitem{cmc-b90}
Again this is required so that both the Cauchy horizon, and cosmological
event horizon, are located at the same advanced time $v = + \infty$.
\bibitem{cmc-b95}
One may be surprised to see the argument of the exponential in
Eq.~(\ref{cmc-e283}) contains the advanced time coordinate $v$
rather than the radial coordinate $r_{*}$. However, along
a radial geodesic that crosses either horizon, it can be seen
[from Eqs.~(\ref{cmc-e281}) and (\ref{cmc-e282}) or
Eqs.~(\ref{cmc-e284}) and (\ref{cmc-e285})] that $dr/dv = \frac{1}{2}
f$. Integrating this, and using Eq.~(\ref{cmc-e40}), one can
easily show that $r_{*} \sim \frac{1}{2} v + k$ as the horizons
are approached. In Sec.~\ref{cmc-s50} we have, without loss of
generality, set the integration constant $k$ to zero.
%
%
\bibitem{cmc-b100}
F. Mellor and I. Moss, Phys. Rev. D {\bf 41}, 403 (1990).
\bibitem{cmc-b110}
Stability in this context requires that the flux, due to the
perturbations, as measured by an observer crossing the Cauchy
horizon be finite there.
\bibitem{cmc-b120}
S. Chandrasekhar and J.B. Hartle, Proc. R. Soc. London
{\bf A384}, 301 (1982).
\bibitem{cmc-b130}
A detailed and informative account of the method used by 
Chandrasekhar and Hartle can be found in~\cite{cmc-b140}.
\bibitem{cmc-b140}
S. Chandrasekhar, {\it The Mathematical Theory of Black Holes}
(Cambridge University Press, Cambridge, England, 1983).
\bibitem{cmc-b150}
V. de Alfaro and T. Regge, {\it Potential Scattering} 
(North-Holland Press, Amsterdam, 1965).
\bibitem{cmc-b160}
P.R. Brady and E. Poisson, Class. Quantum Grav. {\bf 9}, 121 (1992).
\bibitem{cmc-b170}
W.A. Hiscock, Phys. Lett. {\bf 83A}, 110 (1981).
\bibitem{cmc-b180} 
This solution is generally referred to as the
Reissner-Nordstr\"{o}m-Vaidya-de Sitter solution.
Some details of this spacetime are given in Appendix~\ref{cmc-a20}.
\bibitem{cmc-b190}
E. Poisson and W. Israel, Phys. Rev. D {\bf 41}, 1796 (1990).
\bibitem{cmc-b200}
W. Israel, in {\it Black Hole Physics}, edited by V. de Sabbata and
Z. Zhang (Kluwer Press, Amsterdam, 1992).
\bibitem{cmc-b210}
Of course, like the Reissner-Nordstr\"{o}m case,  one can
only keep putting charge on the black hole up to the point of
extremality.
\bibitem{cmc-b220}
F. Mellor and I. Moss, Class. Quantum Grav. {\bf 9}, L43 (1992).
\bibitem{cmc-b230}
P.R. Brady, D. N\'{u}\~{n}ez and S. Sinha, Phys. Rev. D {\bf 47},
4239 (1993).
\bibitem{cmc-b240}
The details of the Poisson-Israel model are dealt with
elsewhere in these proceedings.
\bibitem{cmc-b250}
A. Ori, Phys. Rev. Lett. {\bf 67}, 789 (1991).
\bibitem{cmc-b260}
By strong we mean that we do not get a functional
form for the luminosity function as we did in Sec.~\ref{cmc-s80},
Eq.~(\ref{cmc-e410}).
\bibitem{cmc-b270}
R.H. Price, Phys. Rev. D {\bf 5}, 2419 (1972); Phys. Rev. D {\bf 5}, 
2439 (1972)
\bibitem{cmc-b280}
J. Bi\v{c}\'{a}k, Gen. Relativ. Gravit. {\bf 3}, 331 (1972).
\bibitem{cmc-b285}
C.M. Chambers and I.G. Moss, Class. Quantum Grav. {\bf 11}, 1034 (1994).
\bibitem{cmc-b290}
A concise, but clear, discussion of the Newman-Penrose
formalism can be found in~\cite{cmc-b140} Chap.~1, p.~40.
It transpires that the NP formalism is particularly well
suited to the Kerr and Kerr-de Sitter spacetimes, with
the perturbation equations showing a high degree of
symmetry. Indeed, the first  separation of the 
perturbation equations in the Kerr spacetime was 
completed by Teukolsky in 1973~\cite{cmc-b300} using
the NP formalism.
\bibitem{cmc-b300}
S.A. Teukolsky, Phys. Rev. Lett. {\bf 29}, 1114 (1972).
\bibitem{cmc-b310}
C.M. Chambers, Ph.D thesis, University of Newcastle Upon Tyne, 1995.
\bibitem{cmc-b320}
Kerr-de Sitter, like all black hole spacetimes, is Type D under
the Petrov classification of spacetimes.
\bibitem{cmc-b330}
For an idea of the amount of algebra required, one is guided
to the comments made by Chandrasekhar in ref.~\cite{cmc-b140},
Chap. 9, p. 530.
\bibitem{cmc-b340}
D. Markovi\'{c} and E. Poisson, Phys. Rev. Lett. {\bf 74}, 1280 (1995).
\bibitem{cmc-b350}
For the standard vacuum states, the vacuum stress-energy 
diverges on the future black hole event horizon if the
vacuum state is choosen so that the vacuum stress-energy is
regular on the cosmological horizons~\cite{cmc-b360} and
vice-versa.
\bibitem{cmc-b360}
W.A. Hiscock, Phys. Rev. D {\bf 39}, 1067 (1989).
\bibitem{cmc-b370}
D. Markovi\'{c} and W.G. Unruh, Phys. Rev. D {\bf 43}, 332 (1991).
\bibitem{cmc-b380}
R. Balbinot and R. Bergamini, Phys. Rev. D {\bf 40}, 372 (1989).
%
%
\bibitem{cmc-b390}
Indeed, we know of many series like this whose successive terms
decrease but whose sums do not converge.
A particularly well known example is $\sum_{1}^{\infty} n^{-1}$.
\bibitem{cmc-b400}
C.M. Chambers (unpublished).
\bibitem{cmc-b410}
C. Gundlach, R.H. Price and J. Pullin, Phys. Rev. D {\bf 49}, 883 (1994);
Phys. Rev. D {\bf 49}, 890 (1994).
\bibitem{cmc-b420}
L.M. Burko and A. Ori, {\it Late-time evolution of non-linear gravitational
collapse}, gr-qc9703067.
\bibitem{cmc-b430}
P.R. Brady and J. D. Smith, Phys. Rev. Lett. {\bf 75}, 1256 (1995).
\bibitem{cmc-b440}
L.M. Burko and A. Ori (unpublished).
\bibitem{cmc-b475}
A. Ori, Phys. Rev. Lett. {\bf 68}, 2117 (1992); Phys. Rev. D {\bf 55},
4860 (1997).
%
%
\bibitem{cmc-b450}
P.R. Brady, C.M. Chambers, W. Krivan and P. Laguna, Phys. Rev. D {\bf 55},
7538 (1997).
\bibitem{cmc-b460}
Goldwirth and T. Piran, Phys. Rev. D {\bf 36}, 3575 (1987).
\bibitem{cmc-b470}
D. Garfinkle, Phys. Rev. D{\bf 51}, 5558 (1995).
\bibitem{cmc-b471}
It is not clarified in ref.~\cite{cmc-b100} which perturbation 
mode exhibits this
behavior since only reference to a well in the effective potential
for polar perturbation is made and it is not stated if this
occurs for the lowest radiatable multipole or for
others. It would not be too difficult to plot the potentials
for each type of perturbation, mode, to ascertain this,
but the constant mode is not important enough to warrant this
in these proceedings.
%
%
\bibitem{cmc-b472}
R. Penrose, in {\it General Relativity, an Einstein Centenary Survey},
edited by S.W. Hawking and W. Israel (Cambridge University Press, Cambridge,
1979).
\bibitem{cmc-b473}
S.P. Trivedi, Phys. Rev. D {\bf 47}, 4233 (1993).
\bibitem{cmc-b474}
P.R. Anderson, W.A. Hiscock and D.J. Loranz, Phys. Rev. Lett. {\bf 74},
4365 (1995).
%
%
\bibitem{cmc-b480}
C. Barrab\`{e}s and W. Israel,  Phys.~Rev.~D {\bf 43}, 1129 (1991).
%
\end{thebibliography}
\end{document}